\documentclass[twocolumn,onecolappendix,times]{aastex631} % publised 2-column, 10pt, single spaced, 2-column appendix
\accepted{for publication in The Astrophysical Journal}
\shorttitle{Polarization in EUV coronal lines} % <44 chars
\shortauthors{Shchukina et~al.}

\usepackage{amsmath}
\usepackage{dsfont}
\usepackage{ gensymb}

\begin{document}

\title{Coronal Magnetometry with EUV Permitted Lines}

\author[0000-0003-3958-9935]{Nataliia G. Shchukina}
%\email{natasha-ext@iac.es}
\affiliation{Instituto de Astrof\'{\i}sica de Canarias, E-38205 La Laguna, Tenerife, Spain}
\affiliation{Dpto.\ de Astrof\'{\i}sica, Universidad de La Laguna, E-38206 La Laguna, Tenerife, Spain}
\affiliation{Main Astronomical Observatory, National Academy of Sciences, 03143 Kyiv, Ukraine}
%
%\author{Rafael Manso Sainz}
%\affiliation{Third Institute of Physics, University of G\"ottingen, Friedrich-Hund-Platz 1, 37077 G\"ottingen, Germany}
%
\author[0000-0001-5131-4139]{Javier Trujillo Bueno}
%\email{jtb@iac.es}
\affiliation{Instituto de Astrof\'{\i}sica de Canarias, E-38205 La Laguna, Tenerife, Spain}
\affiliation{Dpto.\ de Astrof\'{\i}sica, Universidad de La Laguna, E-38206 La Laguna, Tenerife, Spain}
\affiliation{Consejo Superior de Investigaciones Cient\'{\i}ficas, Spain}
\author[0000-0003-2752-7681]{Supriya Hebbur Dayananda}
\affiliation{Instituto de Astrof\'{\i}sica de Canarias, E-38205 La Laguna, Tenerife, Spain}
\affiliation{Dpto.\ de Astrof\'{\i}sica, Universidad de La Laguna, E-38206 La Laguna, Tenerife, Spain}
\author{Rafael Manso Sainz}
\affiliation{Third Institute of Physics, University of G\"ottingen, Friedrich-Hund-Platz 1, 37077 G\"ottingen, Germany}
\author{Andrii V. Sukhorukov}
\affiliation{Instituto de Astrof\'{\i}sica de Canarias, E-38205 La Laguna, Tenerife, Spain}
\affiliation{Dpto.\ de Astrof\'{\i}sica, Universidad de La Laguna, E-38206 La Laguna, Tenerife, Spain}
\affiliation{Main Astronomical Observatory, National Academy of Sciences, 03143 Kyiv, Ukraine}
\correspondingauthor{Nataliia G. Shchukina}
\email{natasha-ext@iac.es}

\begin{abstract} % single paragraph of <250 words.
A major challenge in solar physics is to obtain empirical information on the magnetic field 
of the million-degree plasma of the solar corona. To this end, we need observables of the solar radiation sensitive 
to the coronal magnetic field. The most familiar observables are the polarization 
signals of visible and near-infrared forbidden lines of highly 
ionized species and some ultraviolet permitted lines, like hydrogen Lyman-$\alpha$.
While the coronal radiation in these spectral lines can only be 
detected for off-limb line of sights, the coronal radiation from permitted 
extreme ultraviolet (EUV) lines can be observed also 
on the solar disk. These coronal lines are mainly collisionally excited, but 
it has been pointed out that some permitted EUV lines can actually be linearly polarized if their 
lower level carries atomic alignment, 
and that their linear polarization is sensitive to the orientation of the coronal magnetic field
\citep[see][]{2009ASPC..405..423M}.
Here we theoretically investigate the linear polarization in permitted EUV lines of a variety of ions:
Fe {\sc x},   Fe {\sc xi},  Fe {\sc xiii},  Fe {\sc xiv}, Si {\sc ix},  and Si {\sc x}.
To this end, we have developed a numerical code, 
which we have applied to investigate the linear polarization and magnetic sensitivity  
of many permitted EUV lines in a one-dimensional
model of the solar corona, providing a list 
of the most promising lines to be further investigated 
for polarimetry with future space telescopes.   
Our next step will be to extend this work by using state-of-the-art three-dimensional coronal models. 
\end{abstract}

%%%%%%%%%%%%%%%%%%%%%%%%%%%%%%%%%%%%%%%%%%%%%%%%%%%%%%%%%% INTRODUCTION
\section{Introduction}
\label{sec:introduction}

If collisions of the coronal ions with electrons and protons are assumed to be isotropic, 
then the ensuing collisional transitions cannot induce {\em directly} 
atomic level polarization (i.e., 
population imbalances 
and quantum coherence between the magnetic sublevels pertaining to the 
$J$-levels of the coronal ions, $J$ being the total angular momentum of the atomic level
under consideration). Under the common assumption of isotropic inelastic collisions 
the only mechanism capable of directly inducing atomic 
level polarization 
is through radiative transitions driven by an anisotropic 
radiation field. At coronal heights the underlying solar disk is 
practically dark at the extreme ultraviolet (EUV) wavelengths of the permitted line transitions 
investigated here, which have wavelengths $\lambda  < 250$ \AA. 
However, the solar disk is very bright at the visible and near-infrared (IR) wavelengths of the 
forbidden transitions that take place 
between the $J$-levels of the ground-term of the coronal ions.
The ensuing anisotropic 
radiation pumping at the forbidden-line wavelengths produces 
atomic polarization in the $J$-levels of the ground terms of the coronal ions. 
This radiatively-induced atomic level 
polarization can be partly transferred to the upper levels of the EUV permitted transitions 
via the isotropic collisions with electrons and protons that excite such upper levels. As a
result, the spontaneously emitted EUV line radiation can be linearly polarized. 
A detailed explanation of this polarization mechanism of EUV coronal lines, with analytical and numerical results, 
can be found in \citet{2009ASPC..405..423M}.

In \citet{2009ASPC..405..423M} it is shown quantitatively  
that the radiation emitted in the permitted Fe {\sc x}
lines at 174.5 \AA\ and 177 \AA\
turns out to be linearly polarized, both for off-limb and on-disk lines of sight. 
First, scattering of continuum
radiation in the forbidden (magnetic dipole) 
6376.29~\AA\ line generates atomic polarization
in the $J=\frac{3}{2}$ level of the ground term $^2P^o$.
Then, collisional transitions pump this radiatively-induced 
ground-level atomic polarization to the excited levels of the coronal ions.
As a result, the spontaneous emission from the upper level of the permitted 
EUV line under consideration can be partially polarized.
The resulting linear polarization is sensitive to the electron density and to the orientation 
of the magnetic field in the observed coronal plasma structure. 
The aim of this paper is to investigate in detail the linear polarization that this mechanism 
introduces in the EUV permitted lines of many highly ionized atoms of the solar corona, 
and to evaluate {their diagnostic potential} 
for coronal magnetic field measurements. In addition to  {Fe {\sc x}},  
we consider the following coronal ions: 
{Fe {\sc xi},  Fe {\sc xiii},   Fe {\sc xiv},  Si {\sc ix},  and  Si {\sc x}}.
  
In this first paper our strategy is the following. 
We consider a {spherically symmetric  
{one-dimensional (1D)} 
model of the quiet solar corona} 
 \citep{2018ApJ...852...52D}, 
which gives the temperature and the electron density as a function of 
height above the solar limb. 
For each height in the model atmosphere we
solve the statistical equilibrium equations for the
elements of the atomic density matrix 
\citep{LL04},
taking into account the 
continuum radiation reaching 
each height from the quiet Sun disk. This allows us to compute the emissivities in the 
various Stokes parameters and to compute the emergent wavelength-integrated Stokes 
signals as a function of height. In this first paper we consider the limit of single scattering 
events in the plane of the sky {(POS)}, while in a next paper we will investigate the impact of the 
integration along the line of sight (LOS).

Some authors have studied the possible 
impact of resonance scattering on some EUV coronal lines 
\citep[e.g.,][]{1984ApJ...285..347W,1999A&A...345..999R,2002A&A...396.1019R,Supriya-Javier-2021,
2024ApJ...971...27K,EUV-pumping}. The wavelengths of the EUV coronal  
lines investigated in this paper are mainly below 250 \AA, and we have assumed that the excitation of their 
upper levels is dominated by inelastic collisions with electrons (i.e., that at the coronal 
heights where the EUV line radiation is emitted by the million-degree plasma 
there is no significant radiative excitation of the ions). Therefore, the linear polarization signals of the EUV permitted 
lines investigated in this paper are only due to the mechanism explained in \citet{2009ASPC..405..423M}. We leave for a 
future investigation the possibility of anisotropic radiation 
pumping at the short EUV line wavelengths considered here. 
For a recent review on magnetic field diagnostics in the solar upper atmosphere see \citet{JTB-TdPA-ARAA}.       

The structure of the paper is as follows.
In Section~\ref{sec:form_problem} we formulate the problem and explain the methods 
of solution we have implemented.
The atmospheric model and the atomic data 
used in the calculations are described in Section~\ref{sec:Input}, and 
the results are presented in Section~\ref{sec:results}.
{Our selection of the most promising EUV lines is presented in 
Section~\ref{sec:selection}}.
Section~\ref{sec:conclusions} summarizes our main conclusions and discusses 
the next steps of our investigation.

%%%%%%%%%%%%%%%%%%%%%%%%%%%%%%%%%%%%%%%%%%%%%%%%%%%%%%%%%% SECTION
%%%%%%%%%%%%%%%%%%%%%%%%%%%%%%%%%%%%%%%%%%%%%%% FORMULATION of the PROBLEM
\section{Formulation of the problem}
\label{sec:form_problem}

We formulate the problem within the framework of the density matrix theory 
of polarization in spectral lines described in \citet{LL04}, which is valid if the 
incident radiation field is flat across a frequency range wider than both the  
Larmor frequency of the ambient magnetic field and the inverse lifetime of the 
levels. As a matter of fact, at the wavelengths of the forbidden line transitions 
the (visible and near-IR) solar-disk   
continuum radiation that illuminates the coronal ions is 
nearly spectrally flat, while at the EUV wavelengths of the permitted line transitions 
the underlying solar disk is practically dark.

% \subsection{The radiation field tensors}
 %\label{sec:rad-tensors}
 
The radiation field is quantified by the 
radiation field tensors $J^K_Q$, with $K=0,1, 2$ and $-K{\le}Q{\le}K$ 
(see equation 5.157 in the 
quoted monograph). We assume that the radiation that
illuminates the coronal ions at any given height above the Sun's visible limb is  
the continuum radiation coming from the underlying solar disk, 
and that it is unpolarized and axially symmetric around the solar radius 
vector through the considered 
spatial point in the corona (hereafter, the local vertical direction). Therefore, 
in the reference system where the Z-axis is along the local vertical direction (hereafter, 
the `vert' reference system) the only non-zero 
{components of the} 
radiation field tensors 
are the mean intensity

\begin{equation}
  [J^0_0]_{\rm vert}={1\over{2}}{\int} {\rm d}{\mu}\,I(\nu, \, \mu),
\label{eq:jv00}  
\end{equation}
and the radiation anisotropy

\begin{equation}
  [J^2_0]_{\rm vert}={1\over{4\sqrt{2}}}\int {\rm d}{\mu}\,(3\mu^2-1)\,I(\nu, \, \mu),
\label{eq:jv20}
\end{equation}
where $\mu={\rm cos}{\theta}$ (with $\theta$ the angle between 
the ray path and the vertical),
$\nu$ is the frequency of the incident radiation beam,
{and $I(\nu, \, \mu)$ the intensity.} 

We calculate the $[J^0_0]_{\rm vert}$ and $[J^2_0]_{\rm vert}$ components 
following section 12.3 of \cite{LL04}, which uses a 
quadratic expansion of the limb darkening law ($N=2$ in their equation 12.31).
To this end, we have used the center-to-limb variation 
of the {\em observed} solar continuum intensity
$I({\lambda}, {\theta}) / I({\lambda}, {0})$
given by \citet{1976asqu.book.....A}
in page 171.
The absolute continuum intensity at the solar disk 
center $I({\lambda}, {0})$ measured between the spectral lines is
taken also from \citet[][see the Table in page 172]{1976asqu.book.....A}.

In the reference system where the Z-axis is along the direction of the ambient magnetic
field (hereafter, the `mag.field'' reference system)
the ensuing radiation field tensors, $[J^K_Q]_{\rm mag.field}$, can be obtained 
from the $[J^K_Q]_{\rm vert}$ ones {by rotation,} 
as indicated by equation (13.14) of \cite{LL04}.
Given that the incident radiation field is assumed to be unpolarized and the only non-zero
$[J^K_Q]_{\rm vert}$ components are $[J^0_0]_{\rm vert}$ and $[J^2_0]_{\rm vert}$, the 
components $[J^1_Q]_{\rm mag.field}=0$. Moreover, it is useful to note that

\begin{equation}
 [J^0_0]_{\rm mag.field}=[J^0_0]_{\rm vert},
\label{eq:j00}
\end{equation}
and
 \begin{equation}
 [J^2_0]_{\rm mag.field}= \frac{1}{2}\,(3{{\rm cos}^{2}}{\theta_B}-1)\,[J^2_0]_{\rm vert},
\label{eq:j20}
\end{equation}
where 
$\theta_B$ 
is the angle between 
the magnetic field vector and the local vertical.

For the coronal ions investigated here 
({Fe {\sc x},   Fe {\sc xi},  Fe {\sc xiii},  Fe {\sc xiv}, Si {\sc ix},  and Si {\sc x}})
we can consider atomic models without hyperfine structure 
described in the Russel-Saunders coupling scheme. Therefore, the 
energy levels are characterized by the quantum 
numbers $\alpha$ (a set of quantum numbers describing the electronic configuration,  
the { total} orbital angular momentum, and the total electronic spin) and $J$ (the level's total angular
momentum). The state of each ($\alpha J$) atomic level 
is quantified by the multipolar components $\rho^K_Q(\alpha J)$ of the 
atomic density matrix, with $K=0,1, ..., 2J$ and $-K{\le}Q{\le}K$ 
\citep[see Table 3.6 in][]{LL04}. In particular, $\rho^0_0(\alpha J)$ is proportional to the 
level's overall population, $\rho^2_0(\alpha J)$ is the alignment coefficient 
that quantifies the degree of population imbalance between magnetic sublevels 
having different absolute value of the magnetic quantum numbers $M$, 
while $\rho^2_Q(\alpha J)$ with $Q{\ne}0$ are linear combinations of the quantum coherences 
between magnetic sublevels whose quantum number $M$ differ by $Q$ 
(hereafter, coherences).   
In order to find the $\rho^K_Q(\alpha J)$ components corresponding 
to each ($\alpha J$)-level we have to solve the statistical
equilibrium equations \citep[see chapter 7 in][]{LL04}, 
which {account for collisional transitions  
(dependent on the local electron density and temperature in the coronal plasma)
and radiative transitions} (the calculation of which requires taking into account the 
anisotropy of the continuum radiation coming from the underlying solar disk).
For the coronal ions 
the separation in frequency units 
between the $J$-levels is much larger 
than their natural width. This and the assumption that the solar corona 
is optically thin at the wavelengths of the considered line transitions imply 
that we can safely ignore any significant 
impact due to quantum mechanical interference between 
the magnetic sublevels pertaining to different $J$-levels. Therefore, the relevant statistical
equilibrium equations are those corresponding to the multilevel atom model described in 
chapter 7 of \cite{LL04}.

The permitted EUV coronal lines considered in this paper are electric dipole (E1) transitions 
between short-lived upper levels $u$ and long-lived lower levels $\ell$. 
As summarized above, the atomic polarization of
the lower level of the line transitions  
is caused by anisotropic continuum
radiation pumping in the  
magnetic dipole (M1) forbidden line transitions, which lie at visible and near infrared (IR) wavelengths, 
while the atomic polarization in the upper level of the EUV lines results from 
the transfer to the upper levels of such lower-level polarization 
via isotropic collisions \citep{2009ASPC..405..423M}.  

The Hanle effect is the magnetic-field-induced modification of
the atomic level polarization \citep[e.g.,][]{Trujillo-Bueno-2001}, 
and the critical magnetic field strength for producing
a sizable impact on the polarization of the 
atomic level of angular momentum $J$ under consideration is\footnote{This 
basic formula results from equating the Zeeman splitting of the level with its 
natural width.}

%>>>>>>>>>>>>>>>>>>>>>>>>>>>>EQUATION>>>>>>>>>>>>>>>>>
%
\begin{equation}
B_{\rm H}= \frac{1.137 \times 10^{-7}}{{t_{\rm life}}\,{g_{J}}},
\label{eq:BH}
\end{equation}
%
%>>>>>>>>>>>>>>>>>>>>>>>>>>>>END EQUATION>>>>>>>>>>>>>
where ${t_{\rm life}}$ is the level's radiative lifetime, in seconds, and 
$g_{J}$ the Land\'{e} factor of the ${J}$-level under consideration.
Approximately, the sensitivity of a spectral line to the Hanle effect occurs for 
magnetic strengths ($B$) between $0.2B_{\rm H}$ and $5B_{\rm H}$. For $B{\gg}5B_{\rm H}$ 
the line is in the saturation regime of the Hanle effect, in which there are no quantum coherences 
in the magnetic field reference frame and the line's linear polarization is sensitive 
only to the magnetic field orientation \citep[see][]{LL04}.    

The following two points are very important for understanding the formulation of the 
problem. 

First, the lower levels of the EUV line transitions investigated here  
have very long lifetimes, because
they belong to the ground term of the considered ion.
As a result, the critical magnetic field strength for the 
Hanle effect in the lower level of the considered EUV lines is  
$B_{\rm H}{\rm <}10^{-5}$ G. Given that the magnetic fields of the $10^6$ K 
plasma of the solar corona are much larger, the lower levels of 
our EUV line transitions are in the saturation regime of the Hanle effect \citep[e.g.,][]{LL04}. 
For this reason, and recalling that the incident radiation field is assumed to be unpolarized, 
in the reference frame in which the quantization axis of 
total angular momentum is parallel to the magnetic field vector (i.e., the 
magnetic field reference frame) there is no quantum coherence between 
the magnetic sublevels of such lower levels (i.e., in the magnetic field reference 
frame the only non-vanishing multipolar components of the lower-level 
atomic density matrix are $\rho^K_0(\alpha J_l)$, and with $K$ even because 
$[J^1_Q]_{\rm vert.}=[J^1_Q]_{\rm mag.field}=0$).

Second, since we are assuming that only isotropic collisions with electrons 
and protons can excite the upper levels of the EUV permitted lines considered here, 
it is clear that in the magnetic field reference 
frame the only possible multipolar components of the upper-level 
atomic density matrix are $\rho^K_0(\alpha_u J_u)$ because, as explained above,  
in the lower levels of the EUV lines there are no coherences in the magnetic field reference frame.  
Without upper-level coherences in the magnetic field reference frame 
there is no Hanle effect in the EUV lines \citep[see section 10.3 of][]{LL04}. 
Nevertheless, the Einstein coefficient $A(\alpha_u J_u \rightarrow \alpha_l J_l)$
for spontaneous emission from the upper to the lower level 
in the EUV lines investigated 
here is of the order of $10^{10}\,{\rm s}^{-1}$, or larger, and the lifetime of the 
line's upper level is 
${t_{\rm life}} = 1/ \sum_{i<u} A(\alpha_u J_u \rightarrow \alpha_i J_i) $.
For this reason, and noting that $g_{J}{\sim}1$, the critical magnetic field for the 
Hanle effect in the upper level of our EUV lines that results from the application of Equation~(\ref{eq:BH})
is $B_{\rm H}>2000$ G. Therefore, even if the above-mentioned coherences were present in the 
upper levels of our EUV lines (e.g., because of a significant anisotropic radiation pumping at the EUV line wavelengths) 
there would be no Hanle effect because the magnetic fields of the $10^6$ K 
coronal plasma are much weaker than 2000 G. 

Because of the two reasons mentioned above,  
in the magnetic field reference frame  
the only multipolar components of the atomic density matrix that quantify 
the state of any $J$ level are $[\rho^K_0(\alpha J)]_{\rm mag.field}$. These 
multipolar components can be found by solving the statistical equilibrium equations 
of the so-called no-coherence case \citep[see section 7.4 of][]{LL04}.
%
%, which are identical to those of the zero-field case when using Equation~(\ref{eq:j20}) for $[J^2_0]$ instead of Equation~(\ref{eq:jv20}). 
%

Since the critical magnetic field for 
the upper-level Hanle effect is much larger 
than the coronal magnetic field strength,
and the lower level is in its Hanle saturation
regime, it is important to note that the linear polarization in the EUV lines 
investigated in this paper is sensitive only 
to the orientation of the coronal magnetic field, but not to its strength. Clearly, the linear 
polarization amplitudes are also sensitive to the plasma electron density 
and temperature.

We do not consider the Stokes $V$ signals because 
they are negligible for the EUV lines of the solar corona. This is 
because for coronal temperatures in the range ${10^5 - 10^6}$~K
the typical Doppler width of the EUV lines of iron ions is at least two or three orders 
of magnitude larger than the Zeeman splitting due to the coronal magnetic fields.
                      
 \subsection{Emission coefficients}
 \label{sec:emis_coef}
 
The expressions of the emission coefficients in the 
$I$, $Q$, and $U$ Stokes parameters of the permitted 
EUV-line radiation emitted along the direction $\vec{\Omega}_o$ 
take a particularly simple form when choosing the 
positive Stokes-$Q$ reference direction along the {\em perpendicular} to the 
projection of the magnetic field vector onto the plane perpendicular to $\vec{\Omega}_o$ \citep[see][]{LL04}: 
\begin{equation}
\varepsilon_I (\nu,  \vec{\Omega}_o) =  
\\
C{\Big [ } 1 +  \frac{1}{2\sqrt{2}}(3 {\cos}^2\Theta - 1)\,w^{(2)}_{J_u J_\ell}\,\sigma^2_0{(\alpha_u J_u} ){\Big ] }\, {\phi(\nu)},
\label{eq:emis_I}
\end{equation}
\begin{equation}
\varepsilon_Q (\nu,  \vec{\Omega}_o) = C \frac{3}{2\sqrt{2}} \,{\sin}^2\Theta \,w^{(2)}_{J_u J_\ell}\,\sigma^2_0{(\alpha_u J_u})\, {\phi(\nu)},  
\label{eq:emis_Q}
\end{equation}
\begin{equation}
\varepsilon_U (\nu,  \vec{\Omega}_o) = 0,
\label{eq:emis_U}
\end{equation}
where
\begin{equation}
   C=\frac{h\nu}{4 \pi} N({\alpha_u J_u})A(\alpha_u J_u \rightarrow \alpha_\ell J_\ell), \nonumber
\label{eq:emis_V}  
\end{equation}
$\Theta$ is the angle between the 
magnetic field vector and the direction $\vec{\Omega}_o$, 
$h$ the Planck constant, $\nu$ the frequency of the transition,
$A(\alpha_u J_u \rightarrow \alpha_\ell J_\ell)$ 
the Einstein coefficient for spontaneous emission  
from the upper to the lower level, ${\phi(\nu)}$
the normalized line profile, and 
$J_u$ and $J_\ell$ the total angular momenta for the upper  and lower levels. 
The population of the upper level is
\begin{equation}
N({\alpha_u J_u})=N(\rm ion)\sqrt{(2{J_u} + 1)}{\rho^0_0({\alpha_u J_u})}, %%\nonumber
\label{eq:pop} 
\end{equation}
with 
$N(\rm ion)$ being the total number of the ions 
of given type
per unit volume. 

The fractional alignment $\sigma^2_0(\alpha_u J_u) $ of the upper level  
entering in Equations~(\ref{eq:emis_I}) and (\ref{eq:emis_Q}) is 
\begin{equation}
\sigma^2_0(\alpha_u J_u) = \rho^2_0(\alpha_u J_u) / \rho^0_0(\alpha_u J_u), 
\label{eq:s20} 
\end{equation}
where the 
density matrix element
$\rho^2_0(\alpha_u J_u)$ {is that 
defined in the magnetic reference frame.}

\subsection{The statistical equilibrium equations}
\label{sec:stat_eq}

Given that we are assuming that at EUV wavelengths there is 
no anisotropic radiation pumping and that isotropic collisions with electrons and protons 
dominate the excitation of the upper levels of the permitted lines in our multilevel atomic models,
there are no coherences in the magnetic field reference frame. Therefore, the non-vanishing multipolar components 
$\rho^{K}_{0}(\alpha J)$ of the atomic density matrix in the magnetic field reference frame
can be found by solving the statistical equilibrium
equations of the no-coherence case explained in section 7.4 of \cite{LL04}.
%
%, which are identical to those of the zero-field case when using Equation~(\ref{eq:j20}) for $J^2_0$ instead of Equation~(\ref{eq:jv20}).
% 
Such equations are the following:

\begin{eqnarray}
&&
{\frac{\rm d}{{\rm d} t}}  \rho^{K}_{0} (\alpha J) = 
\sum_{\alpha_\ell J_\ell K_\ell} \rho^{K_\ell}_{0} (\alpha_\ell J_\ell)\, {{t}}_{\rm A} (\alpha J K, \, \alpha_\ell J_\ell K_\ell)  
\nonumber \\&& \ \ \ \ \ \ \ \ \ \ 
+\sum_{\alpha_u J_uK_u} \rho^{K_u}_{0} (\alpha_u J_u) \, {t}_{\rm E} 
(\alpha J K, \, \alpha_u J_u K_u) 
\nonumber \\&& \ \ \ \ \ \ \ \ \ \ 
+{{t}}_{\rm S} (\alpha J K, \, \alpha_u J_u K_u) \Big] 
\nonumber \\&&
-\sum_{K^{\prime}}  \rho^{K^{\prime}}_0 \,  [{{r}}_{\rm A} (\alpha J K \, K^{\prime} ) 
%\\
%    
+{{r}}_{\rm E} (\alpha J K \, K^{\prime})  
+{{r}}_{\rm S} (\alpha J K \, K^{\prime} )\Big]  
\nonumber \\&& \ \ \ \ \ \ \ \ \ \
+\sum_{\alpha_\ell J_\ell} \sqrt{\frac{ 2J_\ell + 1 }{ 2J + 1 }}C^{(K)}_{\rm I} (\alpha J , \alpha_\ell J_\ell) \,  \rho^K_0(\alpha_\ell J_\ell) 
\nonumber \\&& \ \ \ \ \ \ \ \ \ \
+ \sum_{\alpha_u J_u} \sqrt{\frac{ 2J_u + 1 }{ 2J + 1 }}C^{(K)}_{\rm S} (\alpha J , \alpha_u J_u) \, \rho^K_0(\alpha_u J_u) 
\nonumber \\&& \ \ \ \ \ \ \ \ \ \
- \Big[ \sum_{\alpha_u J_u} C^{(0)}_{\rm I} (\alpha_u J_u, \alpha J) 
% \\
%    
 + \sum_{\alpha_\ell J_\ell} C^{(0)}_{\rm S} (\alpha_\ell  J_\ell , \alpha J) 
 \nonumber \\&& \ \ \ \ \ \ \ \ \ \ \ \ \ \ \ \ \ \ \ \ \ \ \ \ \ \ \ \ \ \ \ \ \ \ \ 
 + D^{(K)}(\alpha J ) \Big] \, \rho^K_0(\alpha J)\,=\,0. 
 \label{eq:rhoKQ_SSTR}
\end{eqnarray}
For dipole transitions ($\Delta J = 0, \pm 1,  0\nrightarrow0$), the  
expressions for the transfer rates due to absorptions 
(${{t}}_{\rm A}$), 
stimulated emissions (${{t}}_{\rm S}$), and spontaneous 
emissions (${{t}}_{\rm E}$), as well as the expressions for 
the relaxation rates due to absorptions (${{r}}_{\rm A}$), 
stimulated emissions (${{r}}_{\rm S}$), and spontaneous 
emissions (${{r}}_{\rm E}$), can be found in equations 
(7.20) of \cite{LL04}. With the exception of ${{t}}_{\rm E}$ 
and ${{r}}_{\rm E}$, all the other transfer and relaxation rates
contain the radiation field tensors $[J^0_0]_{\rm mag.field}$ 
and $[J^2_0]_{\rm mag.field}$ given by Eqs.~(\ref{eq:j00}) and (\ref{eq:j20}), which are the 
only relevant ones in the research problem under consideration. 
In Eq.~(\ref{eq:rhoKQ_SSTR}) the quantities $C^{(K)}_{\rm I}$ and  $C^{(K)}_{\rm S}$ 
are the multipole components of the inelastic and superelastic collisional rates, respectively, 
and $D^{(K)}$ are the depolarizing rates due to elastic collisions \citep[see equations 7.87, 7.89, 7.102, and 7.100 
of][]{LL04}.

As indicated in Equation~(\ref{eq:rhoKQ_SSTR}) we assume statistical equilibrium
(i.e., ${\frac{\rm d}{{\rm d} t}}  \rho^{K}_{0} (\alpha J) = 0$). The determinant of  
this system of statistical equilibrium equations is zero and is the reason why its solution would give
the $\rho^{K}_{0} (\alpha J)$ components up to a multiplicative factor. 
To obtain the solution for the $\rho^{K}_{0} (\alpha J)$ components, we add 
the trace equation 
%%%>>>>>>>>>>>>>>>>>>>>>>>>>>>>>>>>>>>>>>>>>>>>>>>>>>>>>>>>>>>>>>>>>>>>>>>>>>>>>>r>>>>>>>>>>>>>>>EQUATION
\begin{equation}
\sum_{\alpha J} \frac{N({\alpha J})} {N(\rm ion)}= \sum_{\alpha J} \sqrt{(2J + 1)} \, \rho^0_0(\alpha J) = 1
\label{eq:trace}
\end{equation}
%%%>>>>>>>>>>>>>>>>>>>>>>>>>>>>>>>>>>>>>>>>>>>>>>>>>>>>>>>>>>>>>>>>>>>>>>>>>>>>>>>>>>>>>>>>>>>>>END EQUATION

The level populations ${N({\alpha J})} $ needed for calculating the emission coefficients
$\varepsilon_I (\nu,  \Theta) $ and $\varepsilon_Q (\nu,  \Theta) $
were obtained using
%%%>>>>>>>>>>>>>>>>>>>>>>>>>>>>>>>>>>>>>>>>>>>>>>>>>>>>>>>>>>>>>>>>>>>>>>>>>>>>>>>>>>>>>>>>>>>>>EQUATION
\begin{equation}
N({\alpha J})= {\frac{N({\alpha J})}{N(\rm ion)}}  {\frac{N(\rm ion)} {N(\rm el)}} {\frac{N(\rm el)} {N_{\rm H}}} {\frac{N_{\rm H}}{N_{\rm e}}}{N_{\rm e}},
\label{eq:NaJ}  
\end{equation}
%%%>>>>>>>>>>>>>>>>>>>>>>>>>>>>>>>>>>>>>>>>>>>>>>>>>>>>>>>>>>>>>>>>>>>>>>>>>>>>>>>>>>>>>>>>>>>>>END EQUATION
where 
${N(\rm el)} / {N_{\rm H}}$ is the abundance of the element under consideration 
relative to hydrogen
and
${N_{\rm H}} / {N_{\rm e}}$ is the hydrogen abundance relative to the electron density.
The relative level population 
${N({\alpha J})}/{N(\rm ion)}$
is obtained from the solution of Equations~(\ref{eq:rhoKQ_SSTR}) and (\ref{eq:trace}).
The ionization fraction ${N(\rm ion)} / {N(\rm el)}$ was taken from
CHIANTI 
{Version 10
database
\citep{2021ApJ...909...38D}.}
The ratio ${N_{\rm H}} / {N_{\rm e}}$  
is in the range $\sim 0.8 - 0.9$, since
for coronal temperatures hydrogen and helium are fully ionized 
\citep{2018LRSP...15....5D}.
In our study we adopt
a proton-to-electron density ratio of 0.85.
\begin{figure*}[ht!]
   \centering
   \includegraphics[width=0.8\linewidth]{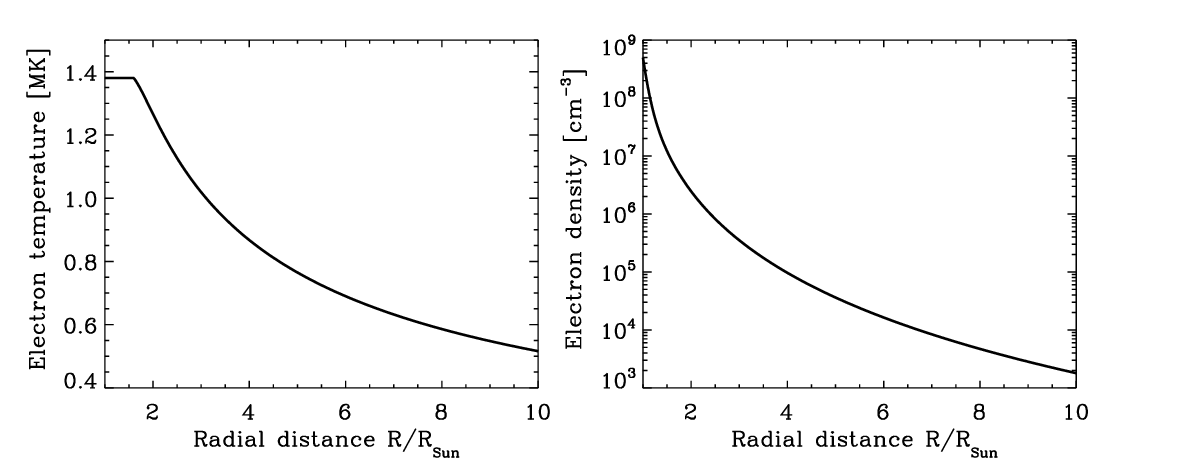}
    \caption{Electron temperature (left panel)  and electron density(right panel)  
  as a function of the radial distance 
  ${R/R_{\rm Sun}}$, 
  {where  $R_{\rm Sun}$ is the solar radius} in 
  spherically symmetric 1D coronal model of the quiet Sun 
 \citep{2018ApJ...852...52D}.
 }
    \label{fig_coronal-model}
\end{figure*}

As explained in \cite{LL04}, the interaction between the ions and the colliders (e.g., electrons)   
is described by a sum of tensor operators of rank $K$ acting on the state vectors of the ion.
Fortunately, in many cases, the interaction is suitably described by just one operator of rank ${\widetilde K}$, 
so that the multipole components $C^{(K)}_{\rm I}$ and $C^{(K)}_{\rm S}$ are 
related to the corresponding multipole components of rank 0 by the following equations 
\citep[see Appendix 4 of][]{LL04}: 

\begin{equation}
    C^{(K)}_{\rm I}  (\alpha J, \alpha_\ell J_\ell) 
  =
   (-1)^{K}
    \frac{
    \begin{Bmatrix}
         J  & J & K\\    
      J_\ell & J_\ell & {\widetilde K}
    \end{Bmatrix}}
    {\begin{Bmatrix}
            J & J & 0 \\
      J_\ell & J_\ell & {\widetilde{K}}
    \end{Bmatrix}}
       C^{(0)}_{\rm I} (\alpha J, \alpha_\ell J_\ell) 
\label{eq:CKI} 
\end{equation}
\begin{equation}
  C^{(K)}_{\rm S}  (\alpha J, \alpha_u J_u)
  =
   (-1)^{K}
    \frac{
    \begin{Bmatrix}
              J   & J  & K \\
      J_u & J_u & {\widetilde K}
    \end{Bmatrix}}
    {\begin{Bmatrix}
      J   & J  & 0 \\
            J_u & J_u & {\widetilde K}
    \end{Bmatrix}}
 C^{(0)}_{\rm S} (\alpha J,  \alpha_u J_u)  ,
\label{eq:CKS} 
\end{equation}
where the curly bracket terms are the Wigner 6-$j$ symbols.
We have ${\widetilde K}=1$ for the E1 transitions and ${\widetilde K}=2$
for the E2 transitions.

Note that in the notation used by 
\citet{1978stat.book.....M} the components of the inelastic $C^{(0)}_{\rm I}$ and superelastic $C^{(0)}_{\rm S}$ 
electron collisional rates are

%%>>>>>>>>>>>>>>>>>>>>>>>>>>>>>>>>>>>>>>>>>>>>>>>>>>>>>>>>>>>>>>>>>>>>>>>>>>>>>>>>>>>>>>>>>>>>>EQUATION
\begin{equation}
C^{(0)}_{\rm I} (\alpha_u J_u, \alpha_\ell J_\ell) = C_{\ell u}
\end{equation}
%%>>>>>>>>>>>>>>>>>>>>>>>>>>>>>>>>>>>>>>>>>>>>>>>>>>>>>>>>>>>>>>>>>>>>>>>>>>>>>>>>>>>>>>>>>>>>>END EQUATION
and
%%>>>>>>>>>>>>>>>>>>>>>>>>>>>>>>>>>>>>>>>>>>>>>>>>>>>>>>>>>>>>>>>>>>>>>>>>>>>>>>>>>>>>>>>>>>>>>EQUATION
\begin{equation}
C^{(0)}_{\rm S} (\alpha_\ell  J_\ell ,  \alpha_u J_u) = C_{u \ell},
\end{equation}
%%%>>>>>>>>>>>>>>>>>>>>>>>>>>>>>>>>>>>>>>>>>>>>>>>>>>>>>>>>>>>>>>>>>>>>>>>>>>>>>>>>>>>>>>>>>>>>>END EQUATION
where $C_{\ell u}$ denotes the inelastic electron collisional rate 
for the transition from the lower level $\ell$ to the upper level $u$,
while $C_{u \ell}$ denotes the superelastic electron collisional rate 
for the transition from the level $u$ to the level $\ell$. 

As in most spectroscopic diagnostics of the solar corona, we assume that the 
electron and ion velocities 
follow a Maxwell-Boltzmann distribution with an electron temperature
${T_{\rm e}}$ 
\citep{2018ApJ...852...52D}. 
In this case
%%>>>>>>>>>>>>>>>>>>>>>>>>>>>>>>>>>>>>>>>>>>>>>>>>>>>>>>>>>>>>>>>>>>>>>>>>>>>>>>>>>>>>>>>>>>>>>EQUATION
%.................................................................................... equation for C_lu from Bely and Regemorter
 \begin{equation}
 C_{{\ell}u}=8.63 \cdot 10^{-6}\,
        {N_{\rm e}}\,
        \frac{\Upsilon_{{\ell}u}(T_{\rm e})}{g_\ell {\sqrt{T_{\rm e}}}}\,
        {\rm exp}\,(-\Delta E_{{\ell}u}/k_B T_{\rm e}),
        \label{CLU_B_R}
\end{equation}
%%>>>>>>>>>>>>>>>>>>>>>>>>>>>>>>>>>>>>>>>>>>>>>>>>>>>>>>>>>>>>>>>>>>>>>>>>>>>>>>>>>>>>>>>>>>>>>END EQUATION
where
${N_{\rm e}}$ is the electron density 
{in $\rm cm^{-3}$, ${T_{\rm e}}$ in K},
$k_B$ is the Boltzmann constant, 
$g_{\ell}=2J_{\ell}+1$ is the statistical weight of the lower level, 
$\Delta E_{{\ell}u}$  is the transition energy,
and the thermally averaged collision strength is
%%%>>>>>>>>>>>>>>>>>>>>>>>>>>>>>>>>>>>>>>>>>>>>>>>>>>>>>>>>>>>>>>>>>>>>>>>>>>>>>>>>>>>>>>>>>>>>>EQUATION
%.................................................................................... equation for Upsilon
\begin{equation}
\Upsilon_{{\ell}u}=\displaystyle\int_0^{\infty} \Omega_{{\ell}u}\exp\left({-\frac{E_u}{{k_B}T_e}}\right)d\left(\frac{E_u}{{k_B}T_e}\right)
\end{equation}
%%>>>>>>>>>>>>>>>>>>>>>>>>>>>>>>>>>>>>>>>>>>>>>>>>>>>>>>>>>>>>>>>>>>>>>>>>>>>>>>>>>>>>>>>>>>>>>END EQUATION
%
with $E_u$  the colliding electron energy after excitation and
$\Omega_{{\ell}u}$ 
the collision strength between levels $\ell$ and $u$ depending on 
$E_{u}/\Delta E_{{\ell}u}$.

For transitions other than E1 and E2
we calculated the multipolar components of the electron collisional rates 
adopting the strong coupling approximation
\citep[e.g.,][]{2006ApJ...651.1229J, 2020SoPh..295...98S}. 

%.............................................................................................. elastic collisional rates

The depolarizing rates $D^{(K)}$ due to elastic collisions 
with neutral hydrogen atoms 
included in Equation~(\ref{eq:rhoKQ_SSTR}) were estimated 
using Equation (7.108) of \citet{LL04}. As expected, it turns out that 
in the quiet solar corona where the neutral hydrogen number density is small
(less then 100 cm$^{-3}$)
the sum of the inelastic and superelastic collisional rate components 
$\Big[ \sum_{\alpha_u J_u} C^{(0)}_{\rm I} (\alpha_u J_u, \alpha J) +
% \\
%    &
 + \sum_{\alpha_\ell J_\ell} C^{(0)}_{\rm S} (\alpha_\ell  J_\ell , \alpha J) \Big] $
for each level 
 ${\alpha J} $
 in Equation~(\ref{eq:rhoKQ_SSTR}) is on average more than eight 
 orders of magnitude larger than the depolarizing elastic rate $D^{(K)}(\alpha J )$.
 Therefore, in all our calculations we neglected such elastic collisions.

Moreover, we did not include inelastic and superelastic collisions with protons,
because our estimations show that, 
for transitions to the highly excited levels of the ions under consideration, 
the proton collisional rates are much smaller  
\citep[see,][for example]{1995ADNDT..60...97B}
than the electron collisional rates given in CHIANTI 
database
\citep{2015A&A...582A..56D, 2021ApJ...909...38D, 2024ApJ...974...71D}.
We also neglected the proton collisions for transition between 
the $J$-levels of the ground terms, 
because they become important only for electron 
temperatures $T_{\rm e} > 3{\times}10^6$~K.

%
%SECTION
%%
\section{The coronal model and the atomic data}
\label{sec:Input}

%CORONA MODEL

\subsection{Coronal model}
\label{sec:corona_model}

We use the spherically symmetric 
{1D} model 
of the quiet-Sun corona proposed by 
\citet{2018ApJ...852...52D}. As seen in the left panel of Figure~\ref{fig_coronal-model}, 
the model's electron temperature is constant 
($ \rm T_e{\approx}1.38{\times}10^{6}$~K) 
for radial distances 
${R/R_{\rm Sun}}\,{\le}\,1.5$ 
while in
higher layers the temperature follows the average of the polar and equator models given in
\citet{2003ApJ...598.1361V}, decreasing till about half a million Kelvin at 
${R/R_{\rm Sun}}\,{=}\,10$.
The right panel of this Figure gives the model's 
electron number density,
as taken 
by \citet{2018ApJ...852...52D}
from \citet{1999JGR...104.9691G},
which goes from $10^9\,{\rm cm}^{-3}$ at the model's surface to about 
$10^3\,{\rm cm}^{-3}$ at 
${R/R_{\rm Sun}}\,{=}\,10$.
At each height in the coronal model we use the corresponding 
electron temperature and electron density
to calculate
the level populations
$N({\alpha J})$ 
from Equations~(\ref{eq:pop}) and  (\ref{eq:NaJ}),
and then apply Equations~(\ref{eq:emis_I}) and (\ref{eq:emis_Q})
to determine
 the Stokes emissivity coefficients
$\varepsilon_I$ and 
$\varepsilon_Q$.
%

%================================================================================== FIGURE 2 Fe X & Fe XI OFF
%
\begin{figure*}
   \centering
      \includegraphics[width=0.8\linewidth]{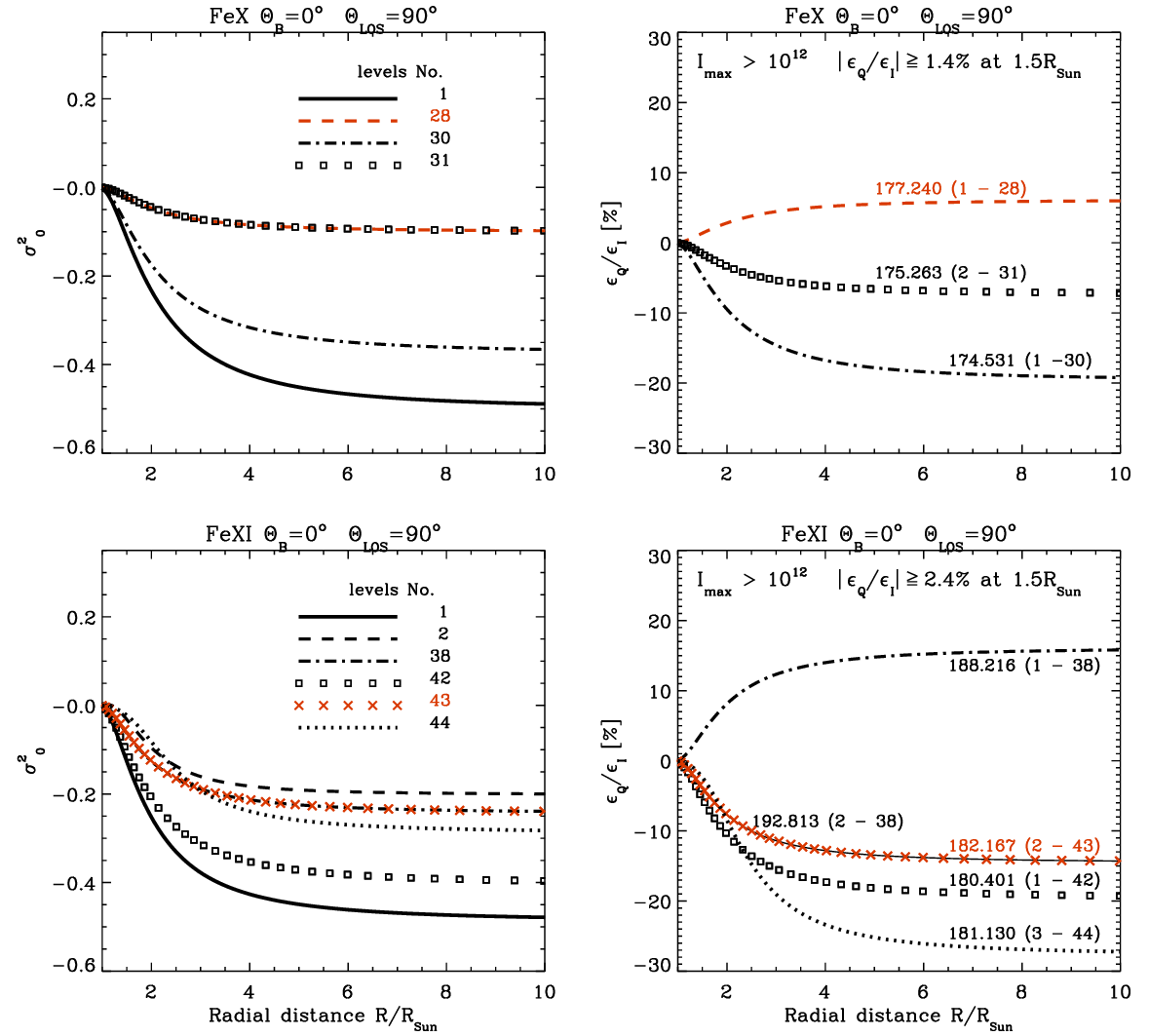}
    \caption{The fractional alignment 
    $\sigma^2_0(\alpha J)$ 
    of the lower and highly excited levels   
                  (left panels)                 
                  and the 
                  ratio of the Stokes $Q$ and $I$ emission coefficients 
                  ${\varepsilon_Q (\nu,  \Theta)}/{\varepsilon_I (\nu,  \Theta)}$
                   (right panels) for selected EUV lines 
                   of the  Fe {\sc x}   
                   and Fe {\sc xi} ions  
                   as a function of the radial distance ${R/R_{\rm Sun}}$ 
                   in the spherically symmetric coronal model of the quiet Sun of
 \cite{2018ApJ...852...52D}.
                The top and bottom panels show the 
                 results for  the  Fe {\sc x}  and  Fe {\sc xi} ions, respectively.
                  The level numbers, wavelengths, and transitions (in brackets) are indicated in the  respective panels.
                  Geometrical configuration ($\Theta_{\rm LOS}$ is the angle between the solar radius vector through the observed point and the LOS):  
off-limb observations 
($\Theta_{\rm LOS} = 90{\degree}$)
of a radial magnetic field
($\theta_{\rm B}=0{\degree}$).
                   For all the \ion{Fe}{10} and \ion{Fe}{11} lines shown,
the  intensity  
   $I_{\rm {max}}$ 
    at  
 ${R/R_{\rm Sun}}\,{=}\,1.0$  
(in units of ${\rm 10^{11}\,photons}\,$  cm$^{-2}$ s$^{-1}$ sr$^{-1}$)
and the 
absolute ratio 
$|\varepsilon_Q/ \varepsilon_I|$  
at a radial distance of 
${R/R_{\rm Sun}=1.5}$ 
are equal to or greater than  the values specified at the top of the right panels.
The positive Stokes-$Q$ direction is the perpendicular to the magnetic field.
}
    \label{fig:s20_QI_FeX_FeXI_OFF}
\end{figure*}
%
%================================================================================== END FIGURE  2
%%%%%%%%%%%%%%%%%%%%%%%%%%%%%%%%%%%%%%%%%%%%%%% Figure 3
\begin{figure*}[ht!]
   \centering
    \includegraphics[width=0.8\linewidth]{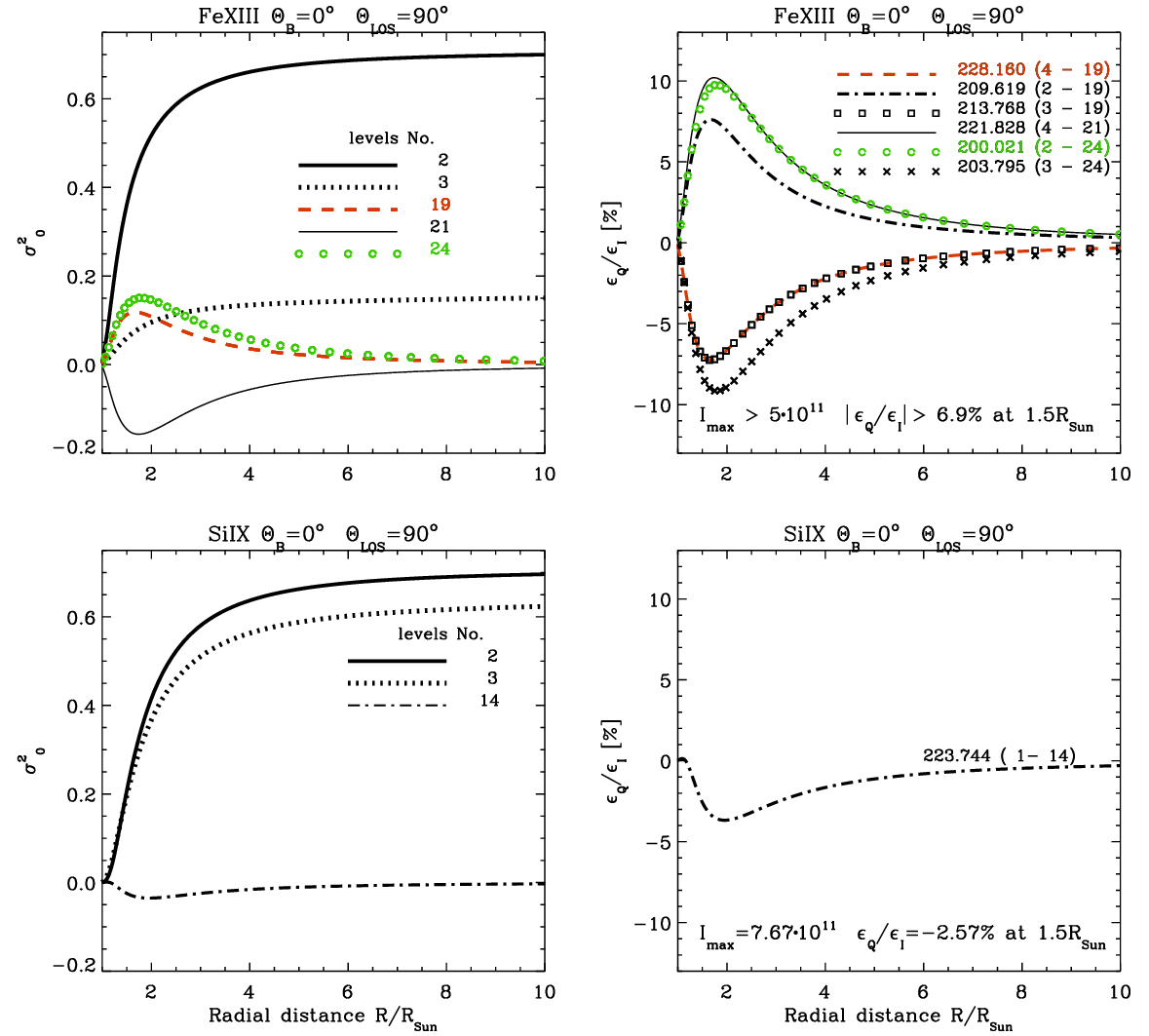}  
    \caption{The fractional alignment $\sigma^2_0(\alpha J)$ of the lower and highly excited levels   
                  (left panels)                 
                  and the 
                  ratio of the Stokes $Q$ and $I$ emission coefficients
            ${\varepsilon_Q (\nu,  \Theta)}/{\varepsilon_I (\nu,  \Theta)}$
                  (right panels) for  selected EUV lines 
                   of the  Fe {\sc xiii} and  Si {\sc ix} ions 
                               as a function of the radial distance ${R/R_{\rm Sun}}$ 
                   in the coronal model of 
 \cite{2018ApJ...852...52D}.
                 The top and bottom panels show the 
                 results for the Fe {\sc xiii} and Si {\sc ix} ions, respectively.
                 The level numbers, wavelengths, and transitions (in brackets) 
                 are indicated in the corresponding panels.
                 Geometrical configuration ($\Theta_{\rm LOS}$ is the angle between the solar radius vector through the observed point and the LOS):  
                off-limb observations 
($\Theta_{\rm LOS} = 90{\degree}$)
of a radial magnetic field
($\theta_{\rm B}=0{\degree}$).
                   For all Fe {\sc xiii} lines shown,  
the intensity 
   $I_{\rm {max}}$ 
    at  
 ${R/R_{\rm Sun}}\,{=}\,1.0$
  (in units of ${\rm 10^{11}\,photons}\,$  cm$^{-2}$ s$^{-1}$ sr$^{-1}$)
and the 
absolute ratio
$|\varepsilon_Q/ \varepsilon_I|$
at a radial distance of 
${R/R_{\rm Sun}=1.5}$ 
are equal to or greater than the values indicated at the bottom of the upper-right panel.
The intensity 
$I_{\rm {max}}$ 
and the ratio
$\varepsilon_Q/ \varepsilon_I$
 at ${R/R_{\rm Sun}=1.5}$
   for the Si {\sc ix} ~223.774~\AA\ line
    are indicated in the bottom section  of the lower-right panel.
    The positive Stokes-$Q$ direction is the perpendicular to the magnetic field.
    }           
    \label{fig:s20_QI_FeXIII_SiIX_OFF}   
\end{figure*}
%
%================================================================================== END FIGURE 3

\subsection{Chemical abundances}
\label{sec:abundances}

For the abundance of iron and silicon we used $A_{\rm Fe} = 7.50$ and $A_{\rm Si} = 7.51$, 
respectively, which are the photospheric values recommended by 
\citet{2009ARA&A..47..481A}. The $A_{\rm Fe}$ value 
agrees with the one obtained by \citet{2001ApJ...550..970S} without 
assuming local thermodynamic equilibrium (LTE), and it is only 
0.04~{dex} larger than the value later reported by 
\citet{2021A&A...653A.141A}.
The $A_{\rm Si}$ value is 0.04~{dex} lower than the non-LTE abundance value determined by
\citet{2012ApJ...755..176S}, but 
it coincides with the value recommended by 
\citet{2021A&A...653A.141A}. 

Photospheric abundances are often used in investigations of the solar corona.
For example, they are used in the CHIANTI database 
\citep{2015A&A...582A..56D, 2018LRSP...15....5D}, 
in studies of quiet and active regions in the solar corona 
\citep[see][and references
therein]{2023ApJ...944..117T},
as well as in simulations of polarized emission in forbidden coronal lines
\citep[e.g.,][]{1999ApJ...522..524C, 2020SoPh..295...98S, 2021SoPh..296..166S, P-CORONA}.

We are aware 
that the chemical composition of the corona is still under debate. 
In particular, the coronal abundances of low 
First Ionization Potential (FIP) elements,  such as Mg, Si, Fe, used in the latest CHIANTI version 11 database
\citep{2024ApJ...974...71D}
are typically higher by a factor of 2 to 4 compared to their photospheric values. 
Moreover, 
the abundances of these chemical elements seem to  
vary substantially in solar regions with different magnetic topologies, such as  
post-flare loops, active  region outflows and coronal mass ejections
\citep[see][and references
therein]{2012ApJ...755...33S, 2015LRSP...12....2L, 2018LRSP...15....5D, 2023ApJ...944..117T,2024ApJ...974...71D}.

%%%%%%%%%%%%%%%%%%%%%%%%%%%%%%%%%%%%%%%%%%%%%%%%%%%%%%%% FIGURE 4
\begin{figure*}[ht!]
   \centering
   \includegraphics[width=0.8\linewidth]{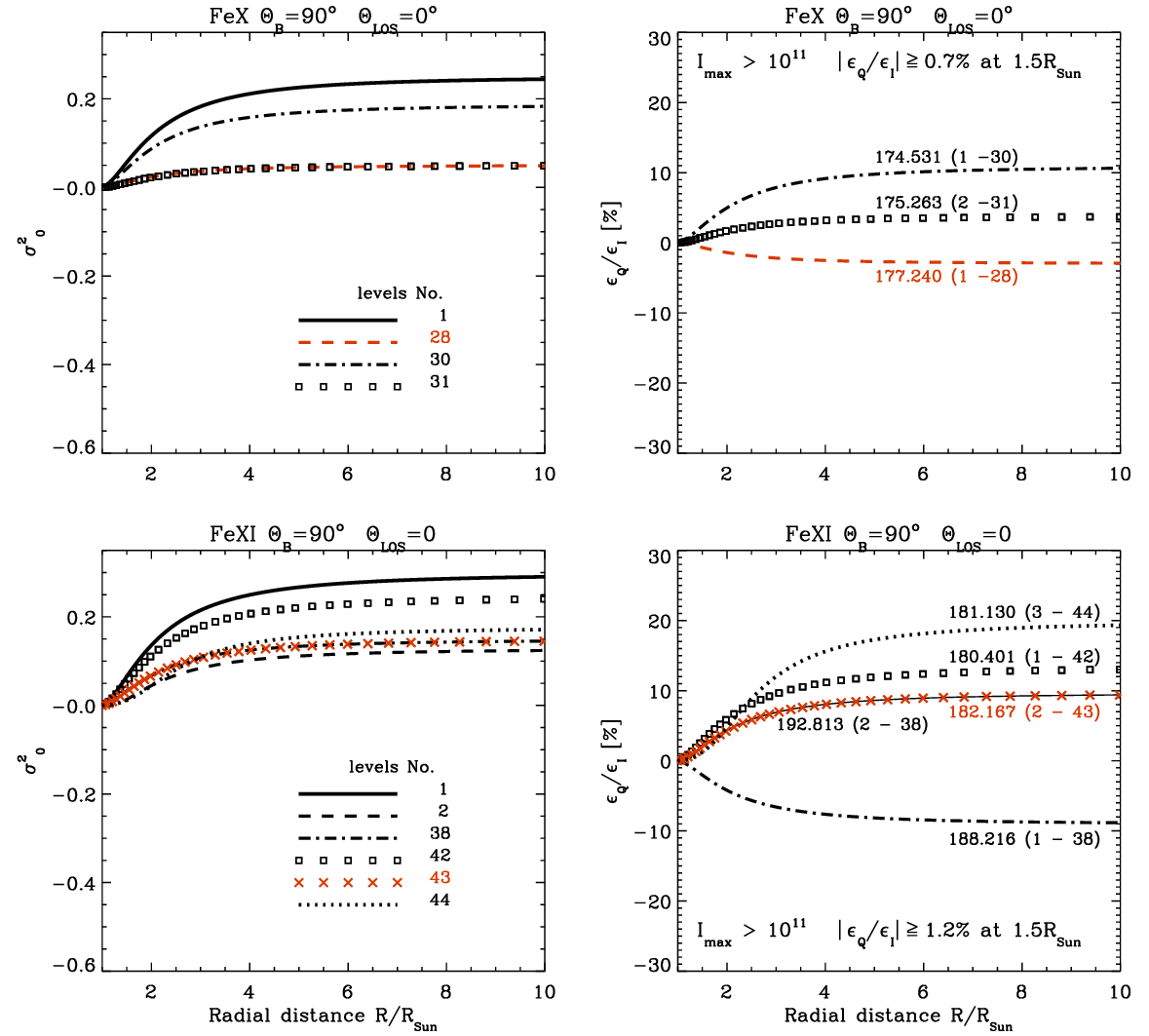}
    \caption{Same as  
    Figure~\ref{fig:s20_QI_FeX_FeXI_OFF}, but for the case of 
 observations 
of a horizontal magnetic field 
                   ($\theta_{\rm B}=90{\degree}$)
   at the disk center ($\Theta_{\rm LOS} = 0{\degree}$).
}
    \label{fig:s20_QI_FeX_FeXI_OVER}
\end{figure*}
%
%================================================================================== END FIGURE 4
%
%%%%%%%%%%%%%%%%%%%%%%%%%%%%%%%%%%%%%%%%%%%%%%%%%%%%%%%%%%%FIGURE 5
\begin{figure*}[ht!]
   \centering
  \includegraphics[width=0.8\linewidth]{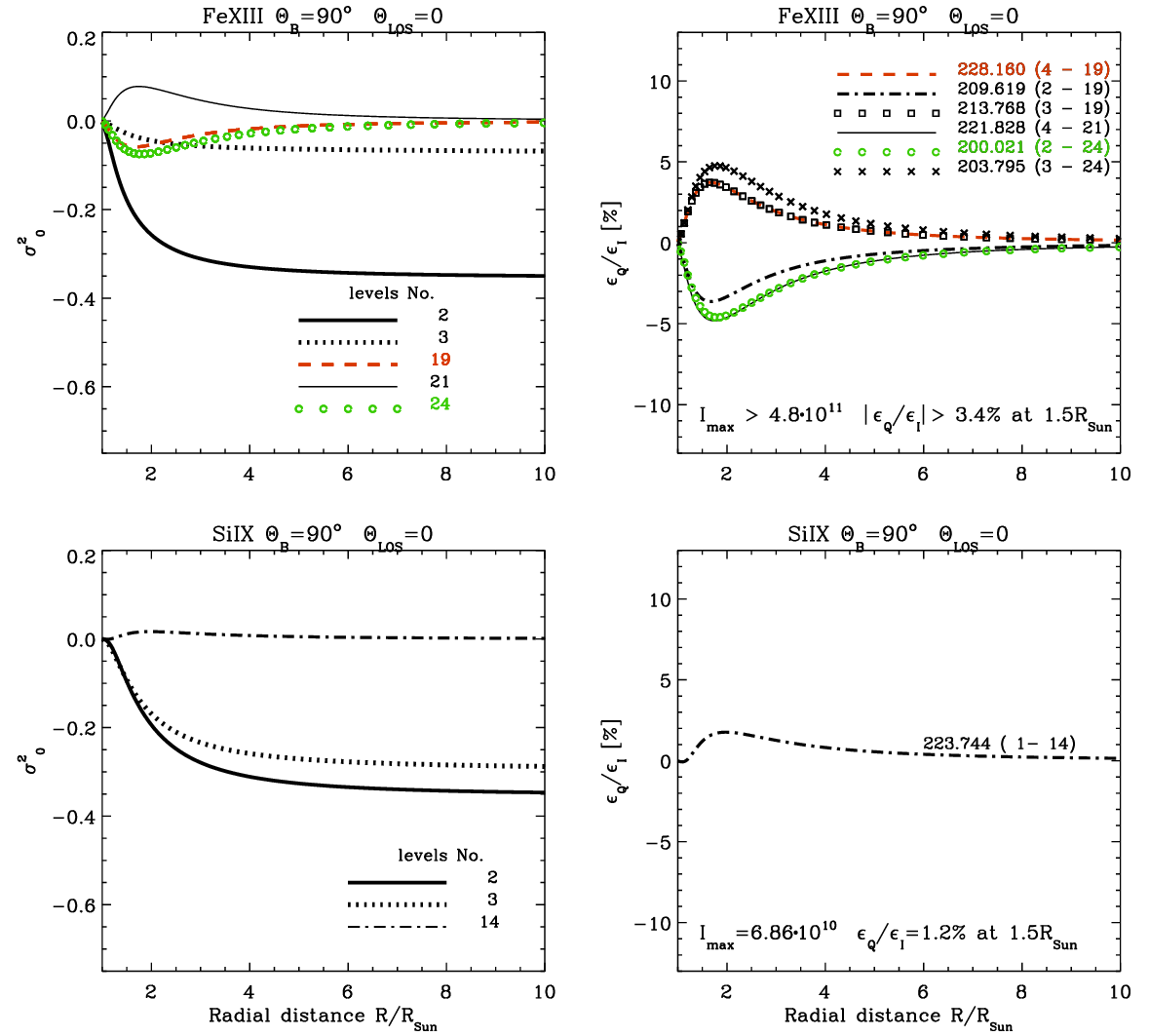}   
    \caption{Same as 
    Figure~\ref{fig:s20_QI_FeXIII_SiIX_OFF}, but for the case of observations of 
 a horizontal magnetic field 
                   ($\theta_{\rm B}=90{\degree}$)
at the disk center ($\Theta_{\rm LOS} = 0{\degree}$).
    }
    \label{fig:s20_QI_FeXIII_SiIX_OVER}
\end{figure*}
%
%%================================================================================== END FIGURE 5

Therefore, {neglecting the FIP effect (i.e. using the smaller photospheric Fe 
and Si abundances) may} result in an underestimation of the Stokes $I$ and $Q$ 
emission coefficients. However, this does not affect 
the fractional atomic alignment $\sigma^2_0$ 
and the fractional linear polarizations $Q/I$ and $U/I$ of the EUV lines under consideration.

%==================================================================== Subsection ATOMIC DATA
\subsection{Atomic data}
\label{sec:atom}

The  atomic data for the Fe {\sc x},   Fe {\sc xi},  Fe {\sc xiii},  Fe {\sc xiv}, Si {\sc ix},  and Si {\sc x} 
ions, including the line identifications, level energies, collision strengths, and
the Einstein coefficients for spontaneous emission, are discussed in detail in several 
papers 
\citep[e.g.,][]{2004A&A...422..731D, 2012A&A...541A..90D, 
2012A&A...537A..38D, 2012A&A...546A..97D, 2015A&A...582A..56D, 
2019ApJS..241...22D, 2019AAS...23431402Y,  2021ApJ...909...38D, 2024ApJ...974...71D}.
In our calculations we used the following model atoms given in CHIANTI {Version 10} database for 
the above-mentioned ions
 \citep{2021ApJ...909...38D}.
\paragraph{Fe {\scriptsize X} ion}
\label{sec:atom_FeX}
The Fe {\sc x}
model atom 
has 552 $J$-levels, 28\,556 radiative bound-bound transitions, and 
152\,076 electron collision bound-bound
transitions.
\paragraph{Fe {\scriptsize XI} ion}
\label{sec:atom_FeXI}
The {Fe {\sc xi}}
model atom 
is the largest one, with 
996 levels,
82\,409 radiative bound-bound transitions, and 
495\,510 electron collision bound-bound
transitions.
\paragraph{Fe {\scriptsize XIII} ion}
\label{sec:atom_FeXIII}
The {Fe {\sc xiii}} 
 model atom  
has 749 $J$-levels,
53\,418 radiative bound-bound transitions, and
280\,126 electron collisional bound-bound transitions.
\paragraph{Fe {\scriptsize XIV} ion}
\label{sec:atom_FeXIV}
The 
{Fe {\sc xiv}}
model atom has 739 $J$-levels,
42\,667 radiative bound-bound 
transitions, and 42\,666 electron collisional bound-bound transitions.
The CHIANTI database for this ion is based mainly on 
the electron collision calculations by \citet{2010ApJS..190..322L}.
At the moment, the electron collision matrix for this ion is incomplete, 
with data only for 1672 transitions.
\paragraph{Si {\scriptsize IX} ion}
\label{sec:atom_SiIX}
{The Si {\sc ix}}
model atom 
{contains
590 levels,
29\,542  radiative bound-bound transitions and 
173\,755 electron collision bound-bound
transitions.}
\paragraph{Si {\scriptsize X} ion}
\label{sec:atom_SiX}
The {Si {\sc x}}
model atom  
has 204 $J$-levels,
5\,017 radiative bound-bound transitions and
20\,706 electron collision bound-bound transitions.

The comprehensive CHIANTI model atoms described above were considered as reference models 
to determine less detailed ones with fewer levels and radiative and collisional transitions.
We concluded that simplified  
{Fe {\sc x},  Fe {\sc xi},  Fe {\sc xiii},  Fe {\sc xiv},  Si {\sc ix}, and Si {\sc x}}
model atoms 
with {the first} 31, 53, 51, 51, 46,  and 25 $J$-levels, respectively, 
would be suitable for our investigation. 
Using such models, the solution of the statistical equilibrium 
Equations~(\ref{eq:rhoKQ_SSTR}) and (\ref{eq:trace})
gives $\rho^K_0(\alpha J)$ {values
with relative errors smaller than} 20\%, {which imply errors smaller than 5\% 
in the fractional alignment $\sigma^2_0$ and in the ratio of emission coefficients
$\varepsilon_Q  / \varepsilon_I$ for the EUV lines considered in this paper. This finding agrees with 
the results obtained by \citet{2025MNRAS.537.3781D} for the {Fe {\sc xiii}} ion.}

\subsection{IDL code}
\label{sec:code}
We have developed an Interactive Data Language (IDL) code
for calculating the intensity and polarization of 
coronal lines based on the formulation of the problem summarized 
in Section~\ref{sec:form_problem}. To facilitate the construction of the atomic model 
needed for the calculations, we have developed also an IDL program 
{ in order to} obtain any desired simplified model atom starting from the comprehensive 
model atoms given in the CHIANTI database
\citep{2015A&A...582A..56D, 2018LRSP...15....5D}.

\section{Results}
\label{sec:results}

We have carried out
calculations for each of the coronal ions mentioned above
(i.e., Fe {\sc x},  Fe {\sc xi},  Fe {\sc xiii},  Fe {\sc xiv},  Si {\sc ix}, and Si {\sc x}) 
using the previously indicated model atoms. 
The calculations have been performed for all  
heights in the coronal model of \citet{2018ApJ...852...52D}
shown in Figure~\ref{fig_coronal-model}, 
between one and ten solar radii.
The larger the height, the larger the 
anisotropy of the continuum radiation coming from the underlying solar disk, but the smaller the 
kinetic temperature (which varies between $1.38{\times}10^6$ K 
and $5{\times}10^5$ K) and the electron density (which varies between $10^9$ 
and $10^3$ ${\rm cm}^{-3}$). 
For the physical conditions corresponding 
to each height in the coronal model we solved the statistical 
equilibrium equations obtaining the $\rho^0_0(\alpha J)$ and $\rho^2_0(\alpha J)$
multipolar components of the atomic density matrix associated to each $J$-level in the 
atomic model under consideration. Then, applying 
Equations~(\ref{eq:s20}), (\ref{eq:emis_I}), and (\ref{eq:emis_Q}) we calculated 
the fractional alignment $\sigma^2_0(\alpha J)$ and the Stokes emission 
coefficients $\epsilon_I$ and $\epsilon_Q$ for any desired magnetic field inclination 
and  LOS.

%%
%%%%%%%%%%%%%%%%%%%%%%%%%%%%%%%%%%%%%%%%%%%%%%%%%%%%%%%%%%%%%%%%%%%%%%%%%%
%%%%%%
%%                To    ADD TABLE 1 Forbidden IR and visible lines
%%
%%TABLE INFRARED and VISIBLE lines producing scattering of the photospheric radiation
\begin{deluxetable}{lrrrlllrr}
\tablecolumns{9}
  \tablewidth{0pc}
\tablecaption{Forbidden lines of the indicated coronal ions.
  }
\label{tab:IR_V_lines}
\decimals
  \tablehead{
 \colhead{Ion} & \colhead{$\lambda$} & \colhead{$\ell$}       & \colhead{$u$}  & \multicolumn{2}{c}{${\rm^{2S+1}L^{o/e}_{J}}$ }       & \colhead{${\rm E}_\ell$}  & \colhead{${\rm E}_u$} & \colhead{$A_{u\ell}$} \\
                       &     \colhead{(\AA)}     &                                 &                         & \colhead{low}                                &    \colhead{up} &  \colhead{(\rm eV)}         & \colhead{(\rm eV)}       &       $\rm s^{-1}$  %%\\
        } 
 \startdata
%%====================== TABLE 1 FeX, Fe XI, Fe XIII, FeIX, Si IX, Si X V & IR lines
%%   
   Fe {\sc x}   &   6376.2 &  1 & 2   &  ${\rm^2P^{o}_{3/2}}$ &  ${\rm^2P^{o}_{1/2}}$   &  0.0  & 1.94 &  69.40   \\
 Fe {\sc xi}     &   7894.0 &  1 & 2   &  ${\rm^3P^{e}_{2}}$   &  ${\rm^3P^{e}_{1}}$    &  0.0    & 1.57 & 43.90   \\
 Fe {\sc xi}    &  61043.0   &  2 & 3   &  ${\rm^3P^{e}_{1}}$   &  ${\rm^3P^{e}_{0}}$    &  1.57  & 1.77 &  0.22   \\
Fe {\sc xiii}    & 10747.0    &  1 & 2   &  ${\rm^3P^{e}_{0}}$   &  ${\rm^3P^{e}_{1}}$    &  0.0    & 1.15 & 14.10   \\
Fe {\sc xiii}     & 10798.0    &  2 & 3   &  ${\rm^3P^{e}_{1}}$   &  ${\rm^3P^{e}_{2}}$    &  1.15  & 2.30 &  9.88   \\
Fe {\sc xiv}    &  5304.4  &  1 & 2   &  ${\rm^2P^{o}_{1/2}}$ &  ${\rm^2P^{o}_{3/2}}$  &  0.0    & 2.33 & 55.20   \\
Si {\sc ix}     & 39277.3  &  1 & 2   &  ${\rm^3P^{e}_{0}}$   &  ${\rm^3P^{e}_{1}}$    &  0.0    & 0.31 &  0.29   \\   
Si {\sc ix}      & 25846.5  &  2 & 3   &  ${\rm^3P^{e}_{1}}$   &  ${\rm^3P^{e}_{2}}$    &  0.31  & 0.79 &  0.77   \\   
Si {\sc x}     & 14305.0    &  1 & 2   &  ${\rm^2P^{o}_{1/2}}$ &  ${\rm^2P^{o}_{3/2}}$  &  0.0    & 0.86 &  3.08   \\   
 \enddata
\tablecomments{The columns give 
 the line wavelengths $\lambda$, the numbers listing the 
 lower ($\ell$) and upper ($u$) levels of the magnetic dipole line transitions, 
 their ${2S+1, L, J}$, and parity ({$o$ or $e$}) quantum numbers, the 
 level energies ${\rm E}_\ell$, ${\rm E}_u$, and the Einstein's coefficients $A_{u\ell}$ for spontaneous emission.}
\end{deluxetable}
%
%%===================================================================================================== TABLE 2 FE X, Fe XI, Fe XIII, Si IX
%
\begin{deluxetable*}{crrccrrrrrrrl}
 \tablecolumns{13}
  \tablewidth{0pc}
  \tablecaption{EUV lines sensitive to the magnetic field orientation.
  }
\label{tab:EUV_polarized_lines_orientation}
\decimals
  \tablehead{          
 \colhead{$\lambda$}   &  \colhead{$\ell$}  & \colhead{$u$} & \multicolumn{2}{c}{${\rm^{2S+1}L^{o/e}_{J}}$ } & \colhead{${\rm E}_\ell$} &  \colhead{${\rm E}_u$} & \colhead{$A_{u\ell}$} &    \colhead{$B_{\rm H}$}  & \colhead{$I_{\rm {max}}$}  & \colhead{$I$}    & \colhead{$\varepsilon_Q / \varepsilon_I$}   & \colhead{Blends}  \\
            &     &   &   \colhead{low}    &    \colhead{up}   &   \colhead{(\rm eV)}    &   \colhead{(\rm eV)}     &      &  \colhead{{\rm (G)}}    & \colhead{1.0${R_{\rm Sun}}$}   & \colhead{1.5${R_{\rm Sun}}$} & \colhead{1.5${R_{\rm Sun}}$}   &  
        } 
 \startdata
               &      &      &                                    &                                    &        &               &  Fe {\sc x}      &              &               &                    &                  &   \\
   177.240 &  1 & 28 &  ${\rm^2P^{o}_{3/2}}$ &  ${\rm^2P^{e}_{3/2}}$ &  0.0 &  69.95    & 154.0& 13409.5 &   125.984   &   0.185  &  $1.368$  & S (Fe {\sc xi}  177.232) \\
  174.531 &  1 & 30 & ${\rm^2P^{o}_{3/2}}$  & ${\rm^2D^{e}_{5/2}}$  &  0.0. &  71.04    & 186.0& 17623.5 &    220.297  &   0.328  &  $-4.692$  &     \\
  175.263  & 2 & 31 & ${\rm^2P^{o}_{1/2}}$  & ${\rm^2D^{e}_{3/2}}$  &  1.94 & 72.69    & 175.0&  25807.0 &    13.380   &   0.010  &   $-1.560$  & B, S (Fe {\sc xi}  175.270) \\
%\hline
               &      &      &                                    &                                   &          &               &  Fe {\sc xi}   &    &           &               &               &   \\
  192.813 &  2 & 38 & ${\rm^3P^{e}_{1}}$  & ${\rm^3P^{o}_{2}}$       &  1.57 & 65.87     &  20.7    & 9016.4  &  40.081    &  0.059& $-3.910$    &  B, Z, S (Fe {\sc xi} 192.817) \\
  188.216 &  1 & 38 & ${\rm^3P^{e}_{2}}$  & ${\rm^3P^{o}_{2}}$       &  0.00 & 65.87     &  96.7    & 9016.4  &  187.111   & 0.270 & $4.015$  & Z \\
  180.401 &  1 & 42 & ${\rm^3P^{e}_{2}}$  & ${\rm^3D^{o}_{3}}$      &   0.00 & 68.73     &  138.0  & 11773.9 &  372.772  &  0.560 & $-5.344$  & Z, S (Fe {\sc x}  180.441) \\
  182.167 &  2 & 43 & ${\rm^3P^{e}_{1}}$  & ${\rm^3D^{o}_{2}}$      &   1.57 & 69.63     &  104.0  & 13132.3 &  54.593   &  0.065 & $-3.836$  &   \\
  178.058 &  1 & 43 & ${\rm^3P^{e}_{2}}$  & ${\rm^3D^{o}_{2}}$      &   0.00 & 69.63     &  28.5    & 13132.3 &  14.953   &  0.017 & $3.936$ &  B \\
  180.594 &  2 & 44 & ${\rm^3P^{e}_{1}}$  & ${\rm^3D^{o}_{1}}$      &   1.57 & 70.22     &  55.7    & 31186.8 &  12.388    &  0.007 & $1.189$  & \\
  181.130 &  3 & 44 & ${\rm^3P^{e}_{0}}$  & ${\rm^3D^{o}_{1}}$      &   1.77 & 70.22     &  78.6    & 31186.8 &  17.484    & 0.010 & $-2.351$  & Z \\
  184.793 &  4 & 45 & ${\rm^1D^{e}_{2}}$  & ${\rm^1D^{o}_{2}}$      &   4.68 & 71.77     &  105.0  & 12541.4 &  5.024    &   0.003  & $-1.572$  & Z, S (Fe {\sc x}  184.828) \\
  192.021 &  2 & 39 & ${\rm^3P^{e}_{1}}$  & ${\rm^3S^{o}_{1}}$       &   1.57 & 66.14    &  29.3     & 6005.6   &  6.077    &   0.006 & $3.271$  & Z \\
  189.123 &  2 & 41 & ${\rm^3P^{e}_{1}}$  & ${\rm^3P^{o}_{1}}$       &   1.57 & 67.13    &  37.3     & 9327.9   &  7.039    &   0.005 & $2.515$  &  B, Z \\
  189.711 &  3 & 41 & ${\rm^3P^{e}_{0}}$  & ${\rm^3P^{o}_{1}}$       &   1.77 & 67.13    &  31.0     & 9327.9  &  5.852     &   0.004 & $-4.907$ &  Z \\
%
%\hline
                &     &     &                                    &                                      &        &              &  Fe {\sc xiii}     &              &              &                    &                  &   \\
   228.160 &  4 & 19 & ${\rm^1D^{e}_{2}}$ & ${\rm^3P^{o}_{2}}$       &  5.96  & 60.30   &  15.80          &  3941.6   &  5.454  &  0.002 & $-6.925$   & \\ 
   209.619 &  2 & 19 & ${\rm^3P^{e}_{1}}$ & ${\rm^3P^{o}_{2}}$       &  1.15  & 60.30   &  18.10          &  3941.6   &  6.235  &   0.002 & $7.260$  &  S (Fe {\sc xi}   209.622) \\ 
  213.768 &  3 & 19 & ${\rm^3P^{e}_{2}}$  & ${\rm^3P^{o}_{2}}$       &   2.30 & 60.30   &  18.10          &  3941.6   &  6.248  &  0.002  & $-6.925$   &  B \\ 
  221.828 &  4 & 21 & ${\rm^1D^{e}_{2}}$  & ${\rm^1D^{o}_{2}}$       &   5.96 & 61.85  &  38.10          &  6527.4    &  7.818  &  0.002  & $9.442$  &   S (Fe {\sc x}  221.813)  \\ 
  200.021 &  2 & 24 & ${\rm^3P^{e}_{1}}$  & ${\rm^3D^{o}_{2}}$       &   1.15 & 63.14  &  25.60.         &  6389.3    &  8.840   &  0.002  & $8.675$  &  B \\   
  203.795 &  3 & 24 & ${\rm^3P^{e}_{2}}$  & ${\rm^3D^{o}_{2}}$       &   2.30 & 63.14  &  34.20          &  6389.3     & 11.834  &  0.003 & $-8.201$   & S (Fe {\sc xiii}  203.826)  \\ 
%   
%    \hline
                &     &      &                                 &                                     &                           &                    &  Si {\sc ix}  &         &             &              &                  &   \\ 
 223.744 &  1 & 14 & ${\rm^3P^{e}_{0}}$  & ${\rm^3S^{o}_{1}}$    &   0.00    & 55.41 &  4.620            &  2390.0  &  7.675  &   0.010   &  $-2.571$   &  S (Fe {\sc xiii}  223.778) \\
\enddata
 \tablecomments{The columns give 
 the line wavelengths $\lambda$ (in \AA), the numbers listing the  
 lower ($\ell$) and upper ($u$) levels of the line transitions, 
 their ${2S+1, L, J}$, and parity ({$o$ or $e$}) quantum numbers,  
 the level energies ${\rm E}_\ell$, ${\rm E}_u$,
the Einstein's coefficients for spontaneous emission,
$A_{u\ell}$ (in units of $\rm 10^9\  s^{-1}$), the 
 critical field strength $B_{\rm H}$, the 
 intensities 
 $I_{\rm {max}}$ at 
 ${R/R_{\rm Sun}}\,{=}\,1.0$, the  
 intensities $I$ 
 at 
 ${R/R_{\rm Sun}}\,{=}\,1.5$  
 (integrated over a path length of
 $\pm 10$ solar radii along the LOS), 
given in units of ${\rm 10^{11}\,photons}\,$  cm$^{-2}$ s$^{-1}$ sr$^{-1}$,
and the ratio of the Stokes emissivity coefficients 
 $\varepsilon_Q (\nu,  \Theta) / \varepsilon_I (\nu,  \Theta) \approx Q/I$ 
 (in \%) at 
${R/R_{\rm Sun}}\,{=}\,1.5$ 
outside the limb
calculated
for radial magnetic field 
($\theta_{\rm B} = 0{\degree}$).
The last column provides information about blends with other lines.
}  
\tablerefs{
The letter B indicates Behring et al. (1976),  
Z refers to Del Zanna (2010, 2012a), and  
S to this study. The symbols in brackets indicate the ion and  
wavelength of the spectral line blends when assuming that the spectral resolution is worse than 
0.05~\AA.
}  
\end{deluxetable*}
%
%
  %%===================================================================================================== TABLE 3 FeX, Fe XI, Fe XIII, Si IX, Si X
 \begin{deluxetable*}{crrccrrlrrrrl}
 \tablecolumns{13} 
  \tablewidth{0pc}
  \tablecaption{EUV lines sensitive to the strength and orientation 
  of the  magnetic field.
  }
\label{tab:EUV_polarized_lines_magnitude}
\decimals
  \tablehead{          
              \colhead{$\lambda$} &  \colhead{$\ell$}  & \colhead{$u$} & \multicolumn{2}{c}{${\rm^{2S+1}L^{o/e}_{J}}$ }  &   \colhead{${\rm E}_\ell$} &  \colhead{${\rm E}_u$} & \colhead{$A_{u\ell}$} &    \colhead{$B_{\rm H}$}   &
              \multicolumn{2}{c} {$0.2B_{\rm{H}}-5B_{\rm{H}}$}  &   \colhead{$I_{\rm {max}}$}   &   \colhead{Blends}  \\              
               \colhead{(\AA)}        &                            &                         &  \colhead{low}    &    \colhead{up}        &      \colhead{(\rm eV)}        &    \colhead{(\rm eV)}     &    &       \colhead{{\rm (G)}}       &         &  \colhead{{\rm (G)}} $\  \  \  \  \  \  \  \  $      & \colhead{1.0${R_{\rm Sun}}$}       &                                                                                                     } 
     \startdata
                      &     &      &                                    &                                    &         &               &  Fe {\sc x}      &              &          &                  &            &  \\         
      225.856 &  1 & 19  &  ${\rm^2P^{o}_{3/2}}$ &  ${\rm^2D^{e}_{5/2}}$ &  0.0  & 54.90     &   0.152               &   14.4    & 2.9    & 72.0          &  6.243 $\  \  $      &  B   \\
                     &     &      &                                    &                                   &         &               &  Fe {\sc xi}      &                &           &                 &              &  \\      
     369.163  &  2  &  6 & ${\rm^3P^{e}_{1}}$     & ${\rm^3P^{o}_{2}}$     & 1.57  & 35.16     &  0.725              &  235.9     &  47.2 & 1179.5       &  20.666 $\  \  $     &    \\   
     352.670  &  1  &  6 & ${\rm^3P^{e}_{2}}$     & ${\rm^3P^{0}_{2}}$     & 0.00  & 35.16     &  2.320              &  235.9     &  47.2 & 1179.5       &  66.088 $\  \  $     &    \\
   356.519   &  2  &  7  & ${\rm^3P^{e}_{1}}$     & ${\rm^3P^{o}_{1}}$     & 1.57  & 36.35     &  0.781              &  249.7     &  49.9   & 1248.5     &    9.002  $\  \  $      &      \\
  358.613    &  3  &  7  & ${\rm^3P^{e}_{0}}$     & ${\rm^3P^{o}_{1}}$     & 1.77  & 36.35     &  0.998              &  249.7     &  49.9   &  1248.5    &  11.513   $\  \  $    &    \\
  341.113    &  1  &  7  & ${\rm^3P^{e}_{2}}$     & ${\rm^3P^{o}_{1}}$     &  0.00 & 36.35     &  1.500               &  249.7     & 49.9   &  1248.5    &   17.295  $\  \  $    &  \\ 
  257.772    &  1  & 12 & ${\rm^3P^{e}_{2}}$     & ${\rm^5D^{o}_{2}}$     &  0.00 & 48.10    &  0.0572              &    4.4      &    0.9   &      22.0    &     8.601  $\  \  $    &    B, Z   \\
  257.554    &  1  & 13 & ${\rm^3P^{e}_{2}}$     & ${\rm^5D^{o}_{3}}$     &  0.00 & 48.14    &  0.0197              &    1.5      &    0.3   &       7.5     &   16.818  $\  \  $    &  Z, S (Fe {\sc xiii}   257.510)  \\
  242.215    &  1  & 15 & ${\rm^3P^{e}_{2}}$     & ${\rm^3D^{o}_{2}}$     &  0.00 & 51.19    &  0.0358              &    5.1      &    1.0   &     25.5     &     6.388   $\  \  $   & Z    \\ 
  240.717    &  1  & 16 & ${\rm^3P^{e}_{2}}$     & ${\rm^3D^{o}_{3}}$     &  0.00 & 51.51    &  0.0018              &    0.2      &    0.04 &     1.0       &    10.406  $\  \  $   &  B, Z, S (Fe {\sc xiii}   240.696) \\ 
  206.258    &  1  & 29 & ${\rm^3P^{e}_{2}}$     & ${\rm^3P^{o}_{1}}$     &  0.00 & 60.11    &  2.410                 &  214.6   &   42.9  &  1073.0     &      5.810  $\  \  $   & Z, S (Fe {\sc xi}  206.268) \\    
  202.424    &  1  & 34 & ${\rm^3P^{e}_{2}}$     & ${\rm^3P^{o}_{2}}$     &  0.00 & 61.25    &  4.660                 &  445.2   &   89.0  &  2226.0     &    13.711  $\  \  $   &  Z, S (Fe {\sc xi}   202.450) \\
  201.112    &  1  & 35 & ${\rm^3P^{e}_{2}}$     & ${\rm^3D^{o}_{3}}$     &  0.00 & 61.65    &  0.449                 &   42.7    &      8.5  &  213.5      &       6.169  $\  \  $  & B, Z, S (Fe {\sc xiii}  201.126) \\  
                  &      &      &                                    &                                      &          &              &  Fe {\sc xiii}       &              &              &                 &              &   \\   
  359.644  &  2   &  8  & ${\rm^3P^{e}_{1}}$    & ${\rm^3D^{o}_{2}}$       &  1.15  & 35.63   &  1.450                 &   147.5  &     29.5  & 737.5       &   13.856 $\  \  $   &  \\ 
                 &     &       &                                    &                                      &             &            & Fe {\sc xiv}       &              &              &                 &               &  \\  
  353.83    &  2 &   7  &   ${\rm^2P^{o}_{3/2}}$ & ${\rm^2D^{e}_{5/2}}$  &  2.34     & 37.38   &  1.970               &   186.7   &   37.3   &   933.5     &   3.773   $\  \ $      &  \\  
                 &     &      &                                     &                                     &             &              &  Si {\sc ix}           &                &            &                &               &   \\  
 345.121   &  2 &  7  & ${\rm^3P^{e}_{1}}$    & ${\rm^3D^{o}_{2}}$       &    0.32   & 36.24   &  2.070                 &   241.3    &    48.3  & 1206.5    &  38.043  $\  \  $     &   \\
 292.759   &  2 & 12 & ${\rm^3P^{e}_{1}}$  & ${\rm^3P^{o}_{2}}$     &    0.32   & 42.67  &  1.390                  &   511.0    &    102.2   &  2555.0    &    10.260 $\  \  $    & S (Si {\sc ix}   292.808) \\
 349.792   &  3 &  7  & ${\rm^3P^{e}_{2}}$  & ${\rm^3D^{o}_{2}}$     &    0.80   & 36.24  &  0.405                  &   241.3   &    48.3    &  1206.5      &      7.466 $\  \  $   &  S (Fe {\sc xi}  349.772) \\
 349.860   &  3 &  9  & ${\rm^3P^{e}_{2}}$  & ${\rm^3D^{o}_{3}}$     &    0.80   & 36.23  &  2.380                  &   203.4   &    40.7     & 1017.0      &    41.112  $\  \  $   &  \\
 296.113  &  3 & 12 & ${\rm^3P^{e}_{2}}$   & ${\rm^3P^{o}_{2}}$     &     0.80   & 42.67  &  5.350                &   511.0     &      102.2  &   2555.0   &    39.506  $\  \  $ &  B \\
 %  \hline  
               &     &      &                                   &                                       &             &            &  Si {\sc x}         &                  &                &     &          & \\
356.037 &  2  &  7  & ${\rm^2P^{o}_{3/2}}$ &  ${\rm^2D^{e}_{5/2}}$   &  0.87     & 35.69 &  2.037                 &  193.0      &   38.6    &   965.0      & 53.629      $\  \  $ &  S (Si {\sc x}    356.049) \\      
       \enddata
 \tablecomments{Similar to Table 2, but for EUV lines that can in principle be 
 sensitive to the magnetic field strength and orientation. 
 The two columns to the right of $B_{\rm H}$
 indicate the approximate range of magnetic sensitivity to the Hanle effect 
 (i.e., $0.2B_{\rm{H}}<B<5B_{\rm{H}}$) of the indicated EUV lines.}        
 \end{deluxetable*}
%
%%%
%%%%%%%%%%%%%%%%%%%%%%%%%%%%%%%%%%%%%%%%%%%%%%% SECTION  RESULTS
%
%
Figures~\ref{fig:s20_QI_FeX_FeXI_OFF} -- \ref{fig:s20_QI_FeXIII_SiIX_OVER} show 
the results { (i.e., $\sigma^2_0$ and $\epsilon_Q/\epsilon_I$) } 
corresponding to the strongest lines of
the following coronal ions:\footnote{We do not show results for the (weaker) permitted EUV 
lines of Fe {\sc ix} and Si {\sc x} because they are similar to those of Fe {\sc xiii} and Si {\sc ix}.}
{ Fe {\sc x}, Fe {\sc xi}, Fe {\sc xiii} and Si {\sc ix}}. The left panels give $\sigma^2_0(\alpha J)$ for some 
of the $J$-levels of the corresponding atomic model and the right panels 
$\varepsilon_Q (\nu,  \Theta) / \varepsilon_I (\nu,  \Theta)$ for some of the 
permitted EUV lines of the coronal ion under consideration, 
as a function of height in the used 
coronal model (up to 10~solar radii).
These results are given for
two different geometrical and magnetic field configurations: 
{ in Figures}~\ref{fig:s20_QI_FeX_FeXI_OFF} --  \ref{fig:s20_QI_FeXIII_SiIX_OFF} 
for off-limb observations 
($\Theta_{\rm LOS} = 90{\degree}$)
of a radial magnetic field
($\theta_{\rm B}=0{\degree}$),
and in Figures~\ref{fig:s20_QI_FeX_FeXI_OVER} -- \ref{fig:s20_QI_FeXIII_SiIX_OVER}
for disk-center observations 
($\Theta_{\rm LOS} = 0{\degree}$)
of a horizontal magnetic field ($\theta_{\rm B}=90{\degree}$). 

Table~\ref{tab:IR_V_lines} gives the list of the infrared and visible
forbidden lines that result from transitions between the 
$J$-levels of the ground term in the considered coronal ions. 
Our results indicate that the scattering of the anisotropic continuum radiation of the solar disk that 
reaches the corona at such forbidden-line wavelengths generates 
atomic polarization in the ground-term $J_\ell$-levels (see the left panels of  
Figures~\ref{fig:s20_QI_FeX_FeXI_OFF} -- \ref{fig:s20_QI_FeXIII_SiIX_OVER}). 
More importantly, the upper levels of many permitted EUV lines 
with Hanle critical magnetic fields $B_{\rm H}>2000$~G 
are also significantly polarized, as a result of the mechanism pointed out by
\citet{2009ASPC..405..423M}: 
the radiatively-induced atomic alignment of the 
ground-term levels is partly transferred to the upper levels of the permitted EUV lines 
via isotropic collisions with electrons. Via spontaneous emissions 
this upper-level atomic alignment produces linear polarization in the emitted EUV line radiation.
The right panels of the figures give a rough estimation of the fractional linear polarization 
wavelength-integrated signals as a function of height above the Sun's visible 
surface.\footnote{This estimation ignores the line of sight integration, because it is just the 
$\epsilon_Q/\epsilon_I$ value in the plane of the sky.}

In order to qualitatively understand the results shown in 
Figures~\ref{fig:s20_QI_FeX_FeXI_OFF} -- \ref{fig:s20_QI_FeXIII_SiIX_OVER}, 
it is convenient to distinguish  
between the following two types of coronal ions: those for which the lowest $J$-level of
the ground term can carry atomic alignment 
(e.g., Fe {\sc x} 
with $J_1=3/2$ and Fe {\sc xi}
with $J_1=2$), and 
those for which the lowest $J$-level cannot carry atomic alignment 
(e.g., Fe {\sc xiii} and Si {\sc ix}
with $J_1=0$, and Fe {\sc xiv} 
and Si {\sc x}
with $J_1=1/2$). As seen in the right 
panels of these figures, 
the variation with the radial distance of the fractional linear polarization signals is different 
for the EUV lines of such two types of ions. The qualitative explanation of this 
different behavior is the following, while a more quantitative one can be found in 
appendices A and B. At any given height in the 
coronal model of Figure~\ref{fig_coronal-model} the most populated level is the lowest-energy level 
of the ion's ground term (i.e., $J_1$), while the first excited level of the ion's ground term (i.e., 
level $J_2$) is significantly less populated, and level $J_3$ is even less populated.
While $N(\alpha_l\,J_1)/N({\rm ion})$ 
turns out to be approximately constant with height
in the solar coronal model, $N(\alpha_l\,J_2)/N({\rm ion})$ decreases with height. 
The $J_1$ level of the ground term in the Fe {\sc x}
and Fe {\sc xi} ions can carry atomic alignment (because $J_1$ 
is neither zero nor $1/2$), while in the Fe {\sc xiii} and Si {\sc ix}
ions (and also in Fe {\sc xiv} and Si {\sc x}) 
the lowest-energy level of the 
ground-term that can carry atomic alignment is $J_2$ 
(see 
Table~\ref{tab:IR_V_lines}).
The wavelengths of the 
forbidden-line transitions between the $J_1$ and $J_2$ levels of the ion's ground term are 
6376.2\AA\ in 
{ Fe {\sc x}},
7894\AA\ in 
{ Fe {\sc xi}},
10747\AA\ in 
{ Fe {\sc xiii}},
5304.4 \AA\ in 
{ Fe {\sc xiv}},
39277.3\AA\ in 
{ Si {\sc ix}},
and 14305\AA\ in 
{ Si {\sc x}}
(see Table~\ref{tab:IR_V_lines}).
At these visible and IR wavelengths the continuum radiation coming from the solar disk  
is significant, and the larger the height in the solar corona the larger the 
anisotropy of such radiation and the larger the fractional atomic alignment of the 
$J_1$ and/or $J_2$ levels. 
As explained in 
Section~\ref{sec:introduction},
the alignment of the upper levels 
of the ion's EUV lines results from the transfer of the atomic alignment in the 
$J$-levels of the ground term via inelastic isotropic collisions { with electrons. Such 
a transfer is clearly greater the larger the
population of the ground-term $J$-level
that carries atomic alignment because of the scattering of anisotropic radiation in the 
forbidden line. }
With these considerations, it is easy to understand the different behavior seen in 
Figures~\ref{fig:s20_QI_FeX_FeXI_OFF} -- \ref{fig:s20_QI_FeXIII_SiIX_OVER} 
for the two types of coronal ions mentioned above.   

\section{Selection of the most useful lines}
\label{sec:selection}

To provide an initial estimate 
of the fractional linear polarization $Q/I$
in {\em off-limb} observations 
(extending up to 2~solar radii) 
the Stokes emission
coefficients $\varepsilon_Q (\nu,  \Theta)$  and $\varepsilon_I (\nu,  \Theta)$  
were integrated over a path length of
$\pm 10$ solar radii 
along the LOS
assuming the 
physical conditions given by the aforementioned 1D model of the quiet Sun corona.
Our estimation indicates that, for the case of  
off-limb observations, 
the fractional linear polarization $Q/I \approx {\varepsilon_Q}/{\varepsilon_I}$
below ${R/R_{\rm Sun}=1.5}$.

From the extensive list of permitted EUV lines 
of the Fe {\sc x},   Fe {\sc xi},  Fe {\sc xiii},  Fe {\sc xiv}, Si {\sc ix},  and Si {\sc x}
ions with non-zero fractional linear polarization $Q/I $, we selected the strongest ones.
In particular, we considered the off-limb case and 
focused on EUV permitted lines 
with wavelength-integrated intensity
$I \geqslant  {5 {\times} 10^{11}}$~photons cm$^{-2}$ s$^{-1}$ sr$^{-1}$
at a radial distance 
${R/R_{\rm Sun}}\,{=}\,1$ 
and absolute linear polarization signal
$\vline  \, Q/I \,\vline \geqslant  1$~\% at a radial distance 
${R/R_{\rm Sun}}\,{=}\,1.5$.
In the 1D coronal model under consideration 
the intensity at the solar limb 
can be regarded as un upper limit for off-limb observations.
For this reason, we denote this intensity by $I_{\rm {max}}$. 

In Table~\ref{tab:EUV_polarized_lines_orientation}
we present the list of EUV lines selected according to the two criteria mentioned above,
including their atomic parameters, the values of their upper-level Hanle  
critical magnetic field strength, 
$B_{\rm H}$, their maximum off-limb intensities, 
$I_{\rm {max}}$, as well as 
their intensity values, $I$, and 
the ratio of the Stokes emission coefficients, 
$\varepsilon_Q (\nu,  \Theta) / \varepsilon_I (\nu,  \Theta)$,
at a radial distance of 
${R/R_{\rm Sun}}\,{=}\,1.5$.

The last column of 
Table~\ref{tab:EUV_polarized_lines_orientation}
indicates that, according to 
\citet{1976ApJ...203..521B}, \citet{2010A&A...514A..41D}, 
and our own identification, some of these EUV lines are blended.
Therefore, their possible { usefulness} for coronal magnetometry needs to be clarified.

We emphasize 
that the intensity $I_{\rm {max}}$ and the ratio $\varepsilon_Q  / \varepsilon_I $ 
presented in 
Table~\ref{tab:EUV_polarized_lines_orientation}
are given only 
for off-limb observations 
($\Theta_{\rm LOS} = 90{\degree}$)
of a radial magnetic field 
($\theta_{\rm B}=0{\degree}$).  
Our calculations indicate that, for observations of a horizontal field 
($\theta_{\rm B}=90{\degree}$)
at  the solar disk  center
($\Theta_{\rm LOS} = 0{\degree}$), 
the absolute values of this ratio for the selected EUV lines 
are approximately half as large.

Finally, we point out that, in addition to the EUV lines listed in 
Table~\ref{tab:EUV_polarized_lines_orientation},
there are other EUV lines of the Fe {\sc x}, Fe {\sc xi}, Fe {\sc xiii}, Fe {\sc xiv}, Si {\sc ix}, and Si {\sc x} ions
that have relatively low values of the
Hanle critical magnetic field strength $B_{\rm H}$ (see Table~\ref{tab:EUV_polarized_lines_magnitude}).
In principle, the linear polarization of these coronal lines   
can be sensitive to the coronal magnetic field strength, because they are not in 
the saturated regime of the Hanle effect. However, for the Hanle effect to operate in these 
EUV permitted lines, upper-level coherences in the magnetic field reference frame must be present. Under the 
assumption of isotropic collisions this would be only possible if there is a significant impact of 
anisotropic radiation pumping at the EUV line wavelengths
\citep[e.g.,][]{EUV-pumping}, which is a possibility that should be carefully 
investigated in the future. 

% 
% CONCLUSIONS
%
\section{Summary \& conclusions}
\label{sec:conclusions}

The EUV radiation from permitted line transitions in highly ionized ions (e.g.,
Fe {\sc x},   Fe {\sc xi},  Fe {\sc xiii},  Fe {\sc xiv}, Si {\sc ix},  and Si {\sc x})
is produced by the million-degree plasma of the solar corona. In the solar 
corona the excitation of the upper levels of such transitions is mainly due 
to inelastic collisions with electrons. At coronal heights the radiation coming from the 
underlying solar disk is very anisotropic, but at EUV wavelengths the intensity of this 
solar-disk radiation is too low so as to produce any significant radiative pumping. However,
as is well known, the radiation coming from the solar disk 
is very significant at the near-IR and visible wavelengths 
of the coronal forbidden lines that result from transitions between the $J$-levels of the 
ion's ground-term, and the ensuing anisotropic radiation pumping induces atomic alignment in
such ground-term levels. \citet{2009ASPC..405..423M} pointed out that this   
radiatively-induced atomic alignment of the ground-term levels can be 
partly transferred to the upper levels of some EUV permitted lines via isotropic inelastic 
collisions with electrons, that the ensuing spontaneous emission produces linear polarization 
in the permitted EUV line under consideration, and that this polarization is sensitive to the
orientation of the magnetic field. In addition to confirming their quantitative results 
for the 
{ Fe {\sc x}}
lines at 174.5 \AA\ and 177 \AA, in this work we 
have investigated the polarization produced by such mechanism in the   
permitted EUV lines of 
{ Fe {\sc x},   Fe {\sc xi},  Fe {\sc xiii},  Fe {\sc xiv}, Si {\sc ix},  and Si {\sc x}},
providing lists of the most interesting lines for further future studies.   

Figures~\ref{fig:s20_QI_FeX_FeXI_OFF} -- \ref{fig:s20_QI_FeXIII_SiIX_OVER} 
show the results obtained for the 
{ Fe {\sc x},   Fe {\sc xi},  Fe {\sc xiii}, and  Si {\sc ix}}
coronal ions,  
at each height in the 1D model of 
Figure~\ref{sec:introduction}.
As seen in these 
figures, the variation with the radial distance of $\varepsilon_Q (\nu,  \Theta) / \varepsilon_I (\nu,  \Theta)$
is different for the permitted EUV lines of 
{ Fe {\sc x}} and { Fe {\sc xi}} compared with those of 
{ Fe {\sc xiii}} and { Si {\sc ix}}. 
As explained in 
Section~\ref{sec:results},
this is because the lowest $J$-level of
the ion's ground term that can carry atomic alignment is $J_1$ 
for 
{ Fe {\sc x}} 
and 
{ Fe {\sc xi}},
while it is $J_2$ for 
{ Fe {\sc xiii}} and { Si {\sc ix}}
($J_1$ being the lowest $J$-level and $J_2$ the first excited $J$-level of the ion's ground term).
The key point here is that the fractional overall population of the $J_1$ level 
(i.e., $N(\alpha_l\,J_1)/N(\rm ion)$) is approximately 
constant with height, while $N(\alpha_l\,J_2)/N(\rm ion)$ decreases with height. 
Consequently, since the alignment of the upper level of the EUV line under consideration 
results from the collisional transfer of the alignment present in the line's lower level, the 
variation with the radial distance of the resulting 
upper-level fractional alignment is different for the two types of 
coronal ions mentioned above. 

The permitted EUV lines given in 
Table~\ref{tab:EUV_polarized_lines_orientation}
are those for which the critical magnetic field for the operation of 
the Hanle effect in their upper-level is larger than 2000 G, reason why their linear 
polarization signals are sensitive to the orientation of the magnetic field, but not to
its strength. The lines given in 
Table~\ref{tab:EUV_polarized_lines_magnitude} 
have much smaller critical Hanle fields and, in principle, 
they can be sensitive to the orientation and the strength of the coronal magnetic field.
However, for the Hanle effect to operate in these EUV lines 
it would be necessary to have radiatively-induced upper-level coherences in the magnetic field reference frame. 

In order to be able to measure
the predicted polarization signals, it is crucial 
to choose the EUV lines with the largest number of emitted solar photons. 
Of the permitted EUV lines we have investigated, 
the most intense EUV lines are the Fe {\sc x}
lines at 
174.531 \AA\ and 177.240 \AA, and the 
Fe {\sc xi} 
lines at 188.216 \AA\ and 180.401 \AA.
However, from these lines the only one that is not blended is the Fe {\sc x}
line at 174.531 \AA.

The next step of our investigation will be to provide predictions from calculations in 
state-of-the-art three-dimensional 
models of the million-degree plasma of coronal structures, along with 
estimations of the exposure times needed for measuring the predicted polarization signals 
assuming reasonable spatial and temporal resolutions. 

\section{Acknowledgments}
\label{sec:acknow}
%\begin{acknowledgments}

We are grateful to Tanaus\'u del Pino Alem\'an (IAC) for suggesting improvements 
to an  earlier version of this paper, as well as to the anonymous referee 
for carefully studying the paper and providing valuable inputs for enhancing the presentation.
We acknowledge financial support from the European Research
Council (ERC) under the European Union's Horizon 2020 research and innovation
program (ERC Advanced Grant agreement No~742265). {NGS} is grateful to the 
Fundaci\'on Occident for funding two working visits at the IAC. JTB 
and SHD acknowledge support from the Agencia Estatal de Investigaci\'on del Ministerio de Ciencia, 
Innovaci\'on y Universidades (MCIU/AEI) under the grant ``Polarimetric Inference of 
Magnetic Fields'' and the European Regional Development Fund (ERDF) with reference
PID2022-136563NB-I00/10.13039/501100011033.

%\end{acknowledgments}

%

\bibliography{ms_FINAL}{}

\begin{thebibliography}{}
\expandafter\ifx\csname natexlab\endcsname\relax\def\natexlab#1{#1}\fi

\bibitem[{{Allen}(1976)}]{1976asqu.book.....A}
{Allen}, C.~W. 1976, {Astrophysical Quantities}, 3rd edn. (University of
  London, Athlone Press)

\bibitem[{{Asplund} {et~al.}(2021){Asplund}, {Amarsi}, \&
  {Grevesse}}]{2021A&A...653A.141A}
{Asplund}, M., {Amarsi}, A.~M., \& {Grevesse}, N. 2021, \aap, 653, A141

\bibitem[{{Asplund} {et~al.}(2009){Asplund}, {Grevesse}, {Sauval}, \&
  {Scott}}]{2009ARA&A..47..481A}
{Asplund}, M., {Grevesse}, N., {Sauval}, A.~J., \& {Scott}, P. 2009, \araa, 47,
  481

\bibitem[{{Behring} {et~al.}(1976){Behring}, {Cohen}, {Feldman}, \&
  {Doschek}}]{1976ApJ...203..521B}
{Behring}, W.~E., {Cohen}, L., {Feldman}, U., \& {Doschek}, G.~A. 1976, \apj,
  203, 521

\bibitem[{{Bhatia} \& {Doschek}(1995)}]{1995ADNDT..60...97B}
{Bhatia}, A.~K., \& {Doschek}, G.~A. 1995, Atomic Data and Nuclear Data Tables,
  60, 97

\bibitem[{{Casini} \& {Judge}(1999)}]{1999ApJ...522..524C}
{Casini}, R., \& {Judge}, P.~G. 1999, \apj, 522, 524

\bibitem[{{Del Zanna}(2010)}]{2010A&A...514A..41D}
{Del Zanna}, G. 2010, \aap, 514, A41

\bibitem[{{Del Zanna}(2012a)}]{2012A&A...537A..38D}
---. 2012a, \aap, 537, A38

\bibitem[{{Del Zanna}(2012b)}]{2012A&A...546A..97D}
---. 2012b, \aap, 546, A97

\bibitem[{{Del Zanna} {et~al.}(2004){Del Zanna}, {Berrington}, \&
  {Mason}}]{2004A&A...422..731D}
{Del Zanna}, G., {Berrington}, K.~A., \& {Mason}, H.~E. 2004, \aap, 422, 731

\bibitem[{{Del Zanna} \& {DeLuca}(2018)}]{2018ApJ...852...52D}
{Del Zanna}, G., \& {DeLuca}, E.~E. 2018, \apj, 852, 52

\bibitem[{{Del Zanna} {et~al.}(2021){Del Zanna}, {Dere}, {Young}, \&
  {Landi}}]{2021ApJ...909...38D}
{Del Zanna}, G., {Dere}, K.~P., {Young}, P.~R., \& {Landi}, E. 2021, \apj, 909,
  38

\bibitem[{{Del Zanna} {et~al.}(2015){Del Zanna}, {Dere}, {Young}, {Landi}, \&
  {Mason}}]{2015A&A...582A..56D}
{Del Zanna}, G., {Dere}, K.~P., {Young}, P.~R., {Landi}, E., \& {Mason}, H.~E.
  2015, \aap, 582, A56

\bibitem[{{Del Zanna} \& {Mason}(2018)}]{2018LRSP...15....5D}
{Del Zanna}, G., \& {Mason}, H.~E. 2018, Living Reviews in Solar Physics, 15, 5

\bibitem[{{Del Zanna} {et~al.}(2012){Del Zanna}, {Storey}, {Badnell}, \&
  {Mason}}]{2012A&A...541A..90D}
{Del Zanna}, G., {Storey}, P.~J., {Badnell}, N.~R., \& {Mason}, H.~E. 2012,
  \aap, 541, A90

\bibitem[{{Del Zanna} \& {Supriya}(2025)}]{2025MNRAS.537.3781D}
{Del Zanna}, G., \& {Supriya}, H.~D. 2025, \mnras, 537, 3781

\bibitem[{{Dere} {et~al.}(2019){Dere}, {Del Zanna}, {Young}, {Landi}, \&
  {Sutherland}}]{2019ApJS..241...22D}
{Dere}, K.~P., {Del Zanna}, G., {Young}, P.~R., {Landi}, E., \& {Sutherland},
  R.~S. 2019, \apjs, 241, 22

\bibitem[{{Dufresne} {et~al.}(2024){Dufresne}, {Del Zanna}, {Young}, {Dere},
  {Deliporanidou}, {Barnes}, \& {Landi}}]{2024ApJ...974...71D}
{Dufresne}, R.~P., {Del Zanna}, G., {Young}, P.~R., {et~al.} 2024, \apj, 974,
  71

\bibitem[{{Gibson} {et~al.}(1999){Gibson}, {Fludra}, {Bagenal}, {Biesecker},
  {del Zanna}, \& {Bromage}}]{1999JGR...104.9691G}
{Gibson}, S.~E., {Fludra}, A., {Bagenal}, F., {et~al.} 1999, \jgr, 104, 9691

\bibitem[{{Judge} {et~al.}(2006){Judge}, {Low}, \&
  {Casini}}]{2006ApJ...651.1229J}
{Judge}, P.~G., {Low}, B.~C., \& {Casini}, R. 2006, \apj, 651, 1229

\bibitem[{{Khan} {et~al.}(2024){Khan}, {Gibson}, {Casini}, \&
  {Nagaraju}}]{2024ApJ...971...27K}
{Khan}, R., {Gibson}, S.~E., {Casini}, R., \& {Nagaraju}, K. 2024, \apj, 971,
  27

\bibitem[{{Laming}(2015)}]{2015LRSP...12....2L}
{Laming}, J.~M. 2015, Living Reviews in Solar Physics, 12, 2

\bibitem[{{Landi Degl'Innocenti} \& {Landolfi}(2004)}]{LL04}
{Landi Degl'Innocenti}, E., \& {Landolfi}, M. 2004, {Polarization in Spectral
  Lines}, Vol. 307 ({Kluwer Academic Publishers}),
  doi:10.1007/978-1-4020-2415-3

\bibitem[{{Liang} {et~al.}(2010){Liang}, {Badnell}, {Crespo L{\'o}pez-Urrutia},
  {Baumann}, {Del Zanna}, {Storey}, {Tawara}, \&
  {Ullrich}}]{2010ApJS..190..322L}
{Liang}, G.~Y., {Badnell}, N.~R., {Crespo L{\'o}pez-Urrutia}, J.~R., {et~al.}
  2010, \apjs, 190, 322

\bibitem[{{Manso Sainz} \& {Trujillo Bueno}(2009)}]{2009ASPC..405..423M}
{Manso Sainz}, R., \& {Trujillo Bueno}, J. 2009, in Astronomical Society of the
  Pacific Conference Series, Vol. 405, Solar Polarization 5: In Honor of Jan
  Stenflo, ed. S.~V. {Berdyugina}, K.~N. {Nagendra}, \& R.~{Ramelli}, 423

\bibitem[{{Mihalas}(1978)}]{1978stat.book.....M}
{Mihalas}, D. 1978, {Stellar atmospheres}, 2nd edn. (W.H. Freeman \& Co.)

\bibitem[{{Raouafi} {et~al.}(1999){Raouafi}, {Lemaire}, \&
  {Sahal-Br{\'e}chot}}]{1999A&A...345..999R}
{Raouafi}, N.~E., {Lemaire}, P., \& {Sahal-Br{\'e}chot}, S. 1999, \aap, 345,
  999

\bibitem[{{Raouafi} {et~al.}(2002){Raouafi}, {Sahal-Br{\'e}chot}, \&
  {Lemaire}}]{2002A&A...396.1019R}
{Raouafi}, N.~E., {Sahal-Br{\'e}chot}, S., \& {Lemaire}, P. 2002, \aap, 396,
  1019

\bibitem[{{Schad} \& {Dima}(2020)}]{2020SoPh..295...98S}
{Schad}, T., \& {Dima}, G. 2020, \solphys, 295, 98

\bibitem[{{Schad} \& {Dima}(2021)}]{2021SoPh..296..166S}
---. 2021, \solphys, 296, 166

\bibitem[{{Schmelz} {et~al.}(2012){Schmelz}, {Reames}, {von Steiger}, \&
  {Basu}}]{2012ApJ...755...33S}
{Schmelz}, J.~T., {Reames}, D.~V., {von Steiger}, R., \& {Basu}, S. 2012, \apj,
  755, 33

\bibitem[{{Seaton} {et~al.}(2025){Seaton}, {Downs}, {Del Zanna}, {West},
  {Thiemann}, {Caspi}, {DeLuca}, {Golub}, {Mason}, {Patel}, {Reeves}, {Rivera},
  \& {Savage}}]{EUV-pumping}
{Seaton}, D.~B., {Downs}, C., {Del Zanna}, G., {et~al.} 2025, arXiv e-prints,
  arXiv:2504.08996

\bibitem[{{Shchukina} {et~al.}(2012){Shchukina}, {Sukhorukov}, \& {Trujillo
  Bueno}}]{2012ApJ...755..176S}
{Shchukina}, N., {Sukhorukov}, A., \& {Trujillo Bueno}, J. 2012, \apj, 755, 176

\bibitem[{{Shchukina} \& {Trujillo Bueno}(2001)}]{2001ApJ...550..970S}
{Shchukina}, N., \& {Trujillo Bueno}, J. 2001, \apj, 550, 970

\bibitem[{{Supriya} {et~al.}(2025){Supriya}, {de Vicente}, {del Pino
  Alem{\'a}n}, {Trujillo Bueno}, \& {Shchukina}}]{P-CORONA}
{Supriya}, H.~D., {de Vicente}, A., {del Pino Alem{\'a}n}, T., {Trujillo
  Bueno}, J., \& {Shchukina}, N. 2025, \apj, in press, (see arXiv:2505.05962)

\bibitem[{{Supriya} {et~al.}(2021){Supriya}, {Trujillo Bueno}, {de Vicente}, \&
  {del Pino Alem{\'a}n}}]{Supriya-Javier-2021}
{Supriya}, H.~D., {Trujillo Bueno}, J., {de Vicente}, A., \& {del Pino
  Alem{\'a}n}, T. 2021, \apj, 920, 140

\bibitem[{{Testa} {et~al.}(2023){Testa}, {Mart{\'\i}nez-Sykora}, \& {De
  Pontieu}}]{2023ApJ...944..117T}
{Testa}, P., {Mart{\'\i}nez-Sykora}, J., \& {De Pontieu}, B. 2023, \apj, 944,
  117

\bibitem[{{Trujillo Bueno}(2001)}]{Trujillo-Bueno-2001}
{Trujillo Bueno}, J. 2001, in Astronomical Society of the Pacific Conference
  Series, Vol. 236, Advanced Solar Polarimetry -- Theory, Observation, and
  Instrumentation, ed. M.~{Sigwarth}, 161

\bibitem[{{Trujillo Bueno} \& {del Pino Alem{\'a}n}(2022)}]{JTB-TdPA-ARAA}
{Trujillo Bueno}, J., \& {del Pino Alem{\'a}n}, T. 2022, \araa, 60, 415

\bibitem[{{V{\'a}squez} {et~al.}(2003){V{\'a}squez}, {van Ballegooijen}, \&
  {Raymond}}]{2003ApJ...598.1361V}
{V{\'a}squez}, A.~M., {van Ballegooijen}, A.~A., \& {Raymond}, J.~C. 2003,
  \apj, 598, 1361

\bibitem[{{Withbroe} \& {Raymond}(1984)}]{1984ApJ...285..347W}
{Withbroe}, G.~L., \& {Raymond}, J.~C. 1984, \apj, 285, 347

\bibitem[{{Young} {et~al.}(2019){Young}, {Dere}, {Del Zanna}, {Landi}, \&
  {Sutherland}}]{2019AAS...23431402Y}
{Young}, P.~R., {Dere}, K.~P., {Del Zanna}, G., {Landi}, E., \& {Sutherland},
  R. 2019, in American Astronomical Society Meeting Abstracts, Vol. 234,
  American Astronomical Society Meeting Abstracts \#234, 314.02

\end{thebibliography}
\bibliographystyle{aasjournal}
\appendix 
In the following two appendices we discuss in more detail the  
atomic polarization of the $J$-levels of the ground term (Appendix A) 
and of the upper levels of the permitted EUV lines studied in this paper (Appendix B).

\section{Atomic polarization of the ground-term levels}
\label{sec:polariz_low_level}

As discussed in Section 4, there are two types of coronal ions in Table 2:
those for which 
the lowest $J$-level of
the ground term can carry atomic alignment 
(e.g., Fe {\sc x}
with $J_1=3/2$ and Fe {\sc xi} 
with $J_1=2$), and 
those for which the lowest $J$-level cannot carry atomic alignment 
(e.g., Fe {\sc xiii}  
and Si {\sc ix}
with $J_1=0$, Fe {\sc xiv} 
and Si {\sc ix} 
with $J_1=1/2$). 

Figures~\ref{fig:s20_QI_FeX_FeXI_OFF} and \ref{fig:s20_QI_FeX_FeXI_OVER}
show results for the first type of ions, while results for the second type are given in 
Figures~\ref{fig:s20_QI_FeXIII_SiIX_OFF} and \ref{fig:s20_QI_FeXIII_SiIX_OVER}.
The wavelengths of the forbidden-line transitions between the ground-term $J$-levels 
are given in Table~\ref{tab:IR_V_lines}. At these visible and IR
forbidden-line wavelengths the solar-disk continuum radiation 
is significant, and the larger the height in the solar corona the larger the 
anisotropy of such radiation and the larger the absolute value of  
the fractional atomic alignment $\sigma^2_0$ of the $J_1$ 
levels of the Fe {\sc x} and Fe {\sc xi} 
ions and of the first excited $J_2$ ground-term levels of the Fe {\sc xiii}
and Si {\sc ix} ions (see left panels of 
Figures~\ref{fig:s20_QI_FeX_FeXI_OFF} -- \ref{fig:s20_QI_FeXIII_SiIX_OVER}).
Additionally, the $J_2$ level (i.e., $^3P^e_{1}$) 
of the Fe {\sc xi} ion also carries atomic polarization,  
due to scattering of the incident anisotropic continuum radiation in the forbidden 61043~\AA\ line,
but its $\sigma^2_0$ absolute value, which 
increases with height, remains smaller.
Likewise, the third excited $J_3$ levels of the Fe {\sc xiii}
and Si {\sc ix} 
ground terms are also polarized by anisotropic radiation pumping in the forbidden-line transitions 
at the 10798~\AA\ and 25846.5~\AA\ infrared wavelengths, 
which occur between the $J_2$ and $J_3$ levels (see Table~\ref{tab:IR_V_lines}).
The fractional alignment 
$\sigma^2_0$ 
of these $J_3$ levels also increases with height, 
but remains lower than that of the $J_2$ levels.
Such a different behavior of the $\sigma^2_0$ values 
for the $J_1$,  $J_2$  and  $J_3$ levels of the ground term 
is easy to understand if one takes into account 
that  the overall population of the 
$J_1$ level is larger than that of level $J_2$, and this larger than that of level $J_3$.

%%=============================================================== FIGURE 6
%
\begin{figure*}[ht!]
   \centering
   \includegraphics[width=0.8\linewidth]{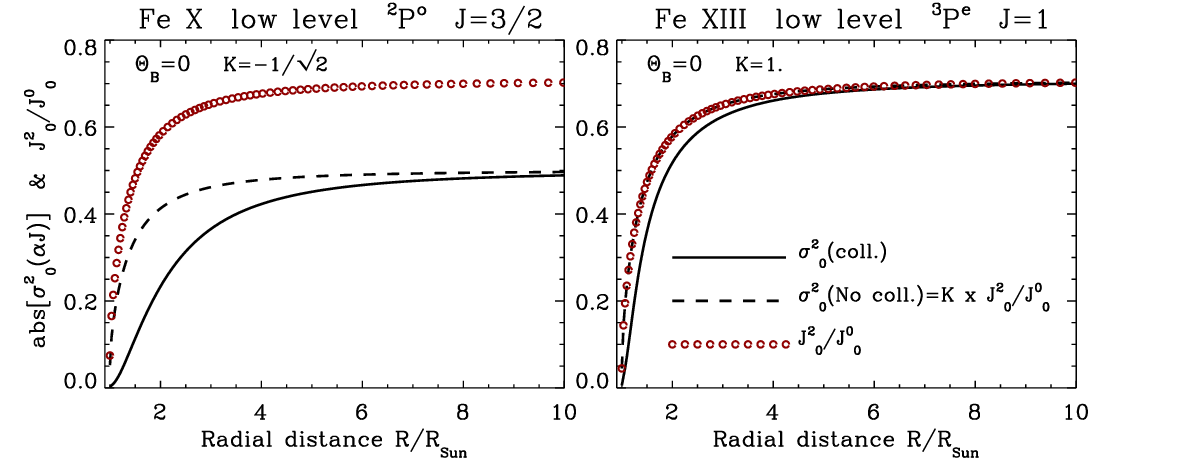}
    \caption{Fractional alignment $\sigma^2_0(\alpha_\ell  J_\ell)$ of  
   the first $^2P^0_{3/2}$ $J_1$-level of the Fe {\sc x} 
   ground term (left panel) and of the second $^3P^e_{1}$ $J_2$-level of the
   Fe {\sc xiii} ground term (right panel) as a function of the radial distance ${R/R_{\rm Sun}}$ in the 
   spherically symmetric 1D quiet-Sun coronal model of Figure 1. 
   Solid lines: $\sigma^2_0(\alpha_\ell  J_\ell)$ values calculated taking into account electron collisions.
 Dashed lines: $\sigma^2_0(\alpha_\ell  J_\ell)$ values obtained in the absence of collisions.
 Red circles: anisotropy of the continuum radiation at 6376~\AA\ (for Fe {\sc x})
 and at 10747~\AA\ (for Fe {\sc xiii}). The results shown are 
 for the case of a vertical magnetic field ($\theta_{\rm B}=0{\degree}$).}
    \label{fig:low_alignment}
\end{figure*}
%
%%=================================================================end  FIG 6

The results presented in Figure~\ref{fig:low_alignment} facilitate a more detailed understanding 
of the atomic polarization behavior of the ground-term levels of the 
{Fe {\sc x},  Fe {\sc xi}, Fe {\sc xiii}, Fe {\sc xiv}, Si {\sc ix}, and Si {\sc x} ions.
This figure shows results only for the 
Fe {\sc x} and Fe {\sc xiii}
ions, because the results for other ions are similar.
Figure~\ref{fig:low_alignment} shows 
that in the absence of electron collisions  
the fractional alignment $\sigma^2_0(\alpha_\ell  J_\ell)$ 
of the lower levels of the ground term 
is proportional to the anisotropy of the incident continuum radiation
%>>>>>>>>>>>>>>>>>>>>>>>>>>>>>>>>>>>>>>>>>>>>>>>>>>>>>>>>>>>>>>>>>>>>>>>>>>>>>>>>>>>>>>>>>>>>>EQUATION
  \begin{equation}
\sigma^2_0(\alpha_\ell  J_\ell) =K \times \frac{J^2_0} {J^0_0} 
\label{eq:s20_low_level} 
 \end{equation}
 %%%>>>>>>>>>>>>>>>>>>>>>>>>>>>>>>>>>>>>>>>>>>>>>>>>>>>>>>>>>>>>>>>>>>>>>>>>>>>>>>>>>>>>>>>>>>>>>END EQUATION
 at 6376~\AA\ (for Fe {\sc x}) 
 and at 10747~\AA\ (for Fe {\sc xiii}).
 
The value of the coefficient $K$ depends 
on which of the levels of the ground term turns out to be polarized.
For the \ion{Fe}{10} and \ion{Fe}{11} ions, whose 
lowest level of the ground term that is aligned
is $J_1$, $K={-1} / {\sqrt{2}}$. 
For the Fe {\sc xiii}, Fe {\sc xiv}, Si {\sc ix}, and Si {\sc x}
ions, the lowest level of the ground term that is aligned is $J_2$. 
In this case, $K=1$ for Fe {\sc xiii}, \ion{Si}{9}, and Si {\sc ix},
while $K={+1} / {\sqrt{2}}$ for Fe {\sc xiv} and Si {\sc ix}. 

Our calculations show that if inelastic collisions with electrons are accounted for,  
the fractional alignment $\sigma^2_0(\alpha_\ell  J_\ell)$
is very well described by formula (6) of
\citet{2009ASPC..405..423M}. The only change is that in the general case 
this formula contains a coefficient $K$ that coincides with ${-1} / {\sqrt{2}}$
only for the Fe {\sc x}
and Fe {\sc xi}
ions. Finally, it should be  emphasized that 
for the case of an inclined magnetic field
the coefficient $K$ has to be multiplied by 
the factor ($3 {\cos}^2\theta_{\rm B} - 1)$.
As a result, for the case of a horizontal magnetic field ($\theta_B=90{\degree}$) 
observed at the solar disk center ($\Theta_{\rm LOS}=0{\degree}$), 
the absolute values of $\sigma^2_0(\alpha_\ell  J_\ell)$ are actually a factor two lower than 
for the case of a radial magnetic field ($\theta_{\rm B}=0{\degree}$) observed off-limb in $90{\degree}$
scattering geometry ($\Theta_{\rm LOS}=90{\degree}$). 
A careful comparison of the $\sigma^2_0(\alpha_\ell  J_\ell)$ values corresponding to these two cases 
(compare Figures~\ref{fig:s20_QI_FeX_FeXI_OFF} and 
~\ref{fig:s20_QI_FeXIII_SiIX_OFF} with  
Figures~\ref{fig:s20_QI_FeX_FeXI_OVER}  and 
\ref{fig:s20_QI_FeXIII_SiIX_OVER}), confirms this conclusion.

Figure~\ref{fig:low_alignment}  
shows that in the 1D coronal model under consideration the inelastic collisions with electrons 
reduce the atomic polarization of the ground-term lower levels.
The electron number density decreases with height and, 
as a result, the lower-level fractional polarization
$\sigma^2_0(\alpha_\ell  J_\ell)$ tends to the maximum value corresponding to 
Equation~(\ref{eq:s20_low_level}).

%
% Appendix B
%

\section{Atomic polarization of the upper levels of the EUV lines}
\label{sec:polariz_high_level}

A comparison of 
Figures~\ref{fig:s20_QI_FeX_FeXI_OFF} and 
~\ref{fig:s20_QI_FeXIII_SiIX_OFF} with 
Figures~\ref{fig:s20_QI_FeX_FeXI_OVER} and 
\ref{fig:s20_QI_FeXIII_SiIX_OVER}
reveals that for the EUV lines with $B_{\rm H}{>}2000$ G, 
the fractional alignment $\sigma^2_0(\alpha_u J_u)$ of their upper levels
and the ratio of their Stokes emissivity coefficients 
$\varepsilon_Q / \varepsilon_I$ show two types of variations 
with the radial distance ${R / R_{\rm Sun}}$.

As shown in 
Figures~\ref{fig:s20_QI_FeX_FeXI_OFF} 
and \ref{fig:s20_QI_FeX_FeXI_OVER}
the variation with the radial distance of the absolute values of 
$\sigma^2_0(\alpha_u J_u)$ and 
$\varepsilon_Q/ \varepsilon_I$  
for the Fe {\sc x} and Fe {\sc xi} 
EUV lines initially increases rapidly, 
then the rate of increase slows down and, 
finally, starting at ${R / R_{\rm Sun}} \approx 6$, the $\sigma^2_0(\alpha_u J_u)$ and 
$\varepsilon_Q/ \varepsilon_I$ values remain constant.

In contrast to the aforementioned EUV lines, 
the height dependence of such quantities for 
the Fe {\sc xiii} and Si {\sc ix} 
EUV lines is different 
(see Figures~\ref{fig:s20_QI_FeXIII_SiIX_OFF} and \ref{fig:s20_QI_FeXIII_SiIX_OVER}). Initially, 
the absolute values of $\sigma^2_0(\alpha_u J_u)$ 
and $\varepsilon_Q/ \varepsilon_I$ also increase with height; 
however, at distances around ${R / R_{\rm Sun}} \approx 1.5 - 3$, 
these quantities reach a maximum value before subsequently decreasing.
A similar behavior is found for the EUV lines of the Fe {\sc xiv} and Si {\sc x} ions.

%
%%=============================================================== FIGURE 7
%%================================================================ to APPENDIX
%s
\begin{figure*}[ht!]
   \centering
      \includegraphics[width=0.8\linewidth]{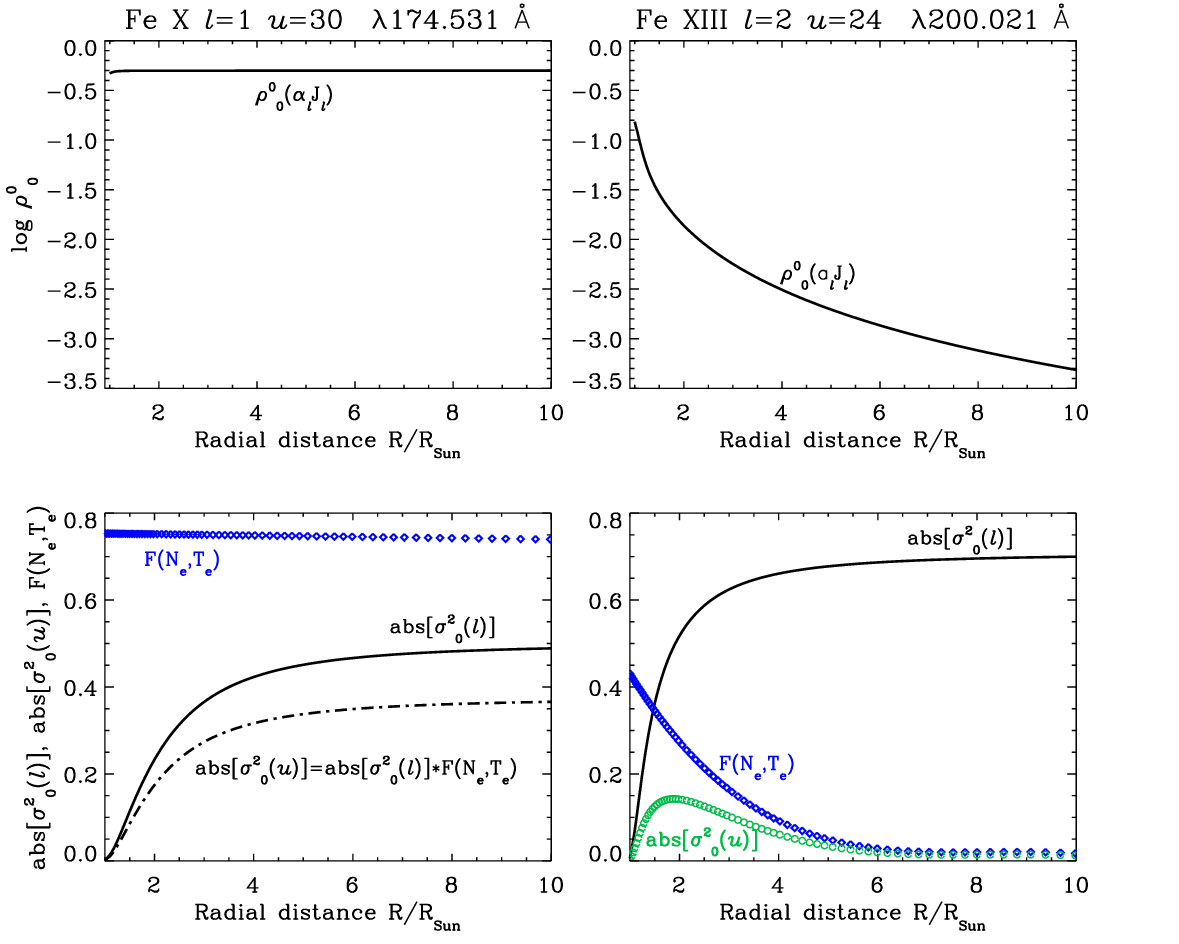}
    \caption{The left and right panels show results for the Fe {\sc x} and Fe {\sc xiii} lines, respectively.
    Top panels: the density matrix elements $\rho^0_0(\alpha_\ell J_\ell)$. Bottom panels:  
the term $F(N_{\rm e}, T_{\rm e})$, the fractional alignment $\sigma^2_0(\alpha_\ell J_\ell)$ (lower level) and  
the fractional alignment $\sigma^2_0(\alpha_u J_u)$ (upper level) for the Fe {\sc x}~174.531~\AA\   
and Fe {\sc xiii}~202.021~\AA\ lines as a function of the radial distance ${R/R_{\rm Sun}}$ in the 
spherically symmetric 1D quiet-Sun coronal model of Figure 1.}
    \label{fig:upper_alignment}
\end{figure*}

%
%%=============================================END FIGURE 7
%

In order to explain this different behavior of the EUV lines 
in such two groups of coronal ions, 
we approximated the fractional alignment $\sigma^2_0(\alpha_u J_u)$ 
of their upper levels using the following expression:
%
%%>>>>>>>>>>>>>>>>>>>>>>>>>>>>>>>>>>>>>>>>>>>>>>>>>>EQUATION
\begin{equation}
\sigma^2_0(\alpha_u J_u) = \sigma^2_0(\alpha_\ell J_\ell)  \times F(N_{\rm e}, T_{\rm e})
\label{eq:s20_approx2} 
\end{equation}
%%%>>>>>>>>>>>>>>>>>>>>>>>>>>>>>>>>>>>>>>>>>>>>>END EQUATION
%
where 
$\sigma^2_0(\alpha_\ell  J_\ell)$
is the 
the fractional alignment of the lower level of the line under consideration,
and 
$F(N_{\rm e}, T_{\rm e})$ 
is a function that depends on 
the coronal electron density $N_{\rm e}$ and the kinetic temperature $T_{\rm e}$.
This expression provides a very good approximation to the results  
obtained with Equation~(\ref{eq:s20}) for $\sigma^2_0(\alpha_u J_u)$ when using 
the $\rho^2_0$ and $\rho^0_0$ values that result from the solution of the statistical equilibrium 
Equations~(\ref{eq:rhoKQ_SSTR}) and (\ref{eq:trace}).
The linear Pearson correlation coefficient between the approximate and exact 
$\sigma^2_0(\alpha_u J_u)$ values exceeds 0.99.

The function 
$F(N_{\rm e}, T_{\rm e})$ 
primarily depends on the density matrix element 
%
%%%>>>>>>>>>>>>>>>>>>>>>>>>>>>>>>>>>>>>>>>>>>>>>>>>>>EQUATION
\begin{equation}
\rho^0_0(\alpha_\ell J_\ell)= \frac{N({\alpha_\ell J_\ell})}{N(\rm ion)} \times  \frac{1}{\sqrt{(2J_\ell + 1)}}, 
\label{eq:r00_low} 
\end{equation}
%%%>>>>>>>>>>>>>>>>>>>>>>>>>>>>>>>>>>>>>>>>>>>>>>>>>>>END EQUATION
%
which in turn is proportional 
to the relative population $N({\alpha_\ell J_\ell}) / N(\rm ion)$ 
of the lower $J_\ell$ level. 
For the case of the Fe {\sc x} 
and Fe {\sc xi} ions the lower level $J_\ell$ is $J_1$ (i.e., the lowest-energy 
level of the ground term).
For the Fe {\sc xiii}, Fe {\sc xiv}, Si {\sc ix}, and  Si {\sc x}}
ions, it is $J_2$ (i.e., the first excited level of the ground term).

Considering that in the solar corona the vast majority of ions  
are in the $J_1$ level, it is expected that the ratio 
$N({\alpha_\ell J_1}) / N(\rm ion)$ 
and, hence, the function $F(N_{\rm e}, T_{\rm e})$ 
will vary only slightly with the radial distance $R/ R_{{\rm Sun}}$.
On the other hand, the ratio 
$N({\alpha_\ell J_2}) / N(\rm ion)$ decreases rapidly with increasing radial distance. 
Consequently, the function $F(N_{\rm e}, T_{\rm e})$ must also decrease with height.

Figure~ \ref{fig:upper_alignment}
illustrates the dependence of
$\rho^0_0(\alpha_\ell J_\ell)$, 
$\sigma^2_0(\alpha_\ell J_\ell)$,
$\sigma^2_0(\alpha_u J_u) $, 
and $F(N_{\rm e}, T_{\rm e})$  
on the radial distance 
${R / R_{\rm Sun}}$,  
for the Fe {\sc x}~174.531~\AA\ 
and Fe {\sc xiii}~202.021~\AA\  lines.
Results for the Fe {\sc x}
and Fe {\sc xiii} 
lines are shown in the left and right panels, respectively.
The top panels of this figure
show that the height dependence of
the density-matrix element $\rho^0_0(\alpha_\ell  J_\ell)$  
for the Fe {\sc x} 
ground-term level $J_1$ 
and for the Fe {\sc xiii} ground-term level $J_2$ 
is very different. As we can see, the values of 
$\rho^0_0(\alpha_\ell  J_1)$  
for the lower level of the 
Fe {\sc x}~174.531~\AA\ line 
changes very little with height, while the values of 
$\rho^0_0(\alpha_\ell  J_2)$ for the lower level of the 
Fe {\sc xiii}~202.021~\AA\ line
decreases by nearly three orders of magnitude.

The bottom panels of Figure~\ref{fig:upper_alignment} show the variation with the radial distance of 
the function $F(N_{\rm e}, T_{\rm e})$   
for both types of ions. Its behavior (see the blue curve) is similar to that of 
$\rho^0_0(\alpha_\ell J_\ell)$. For the Fe {\sc x} ion, $F$ remains nearly constant 
throughout the corona, whereas for the Fe {\sc xiii} ion 
$F$ decreases rapidly with increasing radial distance. As a result,  
the shape of the variation with the radial distance of the absolute value of   
$\sigma^2_0(J_u)$  
for the highly excited upper level $J_{30}$ of the 
Fe {\sc x}~174.531~\AA\ line is similar to that of $\sigma^2_0(J_\ell)$. This is because 
according to Equation B2, the variation with the radial distance of $\sigma^2_0(J_u)$    
follows that of the product of the (nearly constant) function $F$ and 
the (increasing with height) absolute value of the fractional alignment 
$\sigma^2_0(J_\ell)=\sigma^2_0(J_1)$.

The bottom right panel  of 
Figure~\ref{fig:upper_alignment}
shows that for the Fe {\sc xiii}~202.021~\AA\ line 
the shape of the variation with the radial distance of the absolute value of $\sigma^2_0(J_u)$ 
is that given by the green-circles curve, which follows 
that of the product of the (decreasing with height) function $F$ and 
the (increasing with height) absolute value of the fractional alignment 
$\sigma^2_0(J_\ell)=\sigma^2_0(J_2)$.

Finally, we point out that
the above-mentioned results for the EUV lines of  
Fe {\sc x} are similar to those of Fe {\sc xi}. Likewise, the results obtained for the EUV lines of Fe {\sc xiii}
are similar to those of the 
Fe {\sc xiv}, Si {\sc ix}, and Si {\sc x} ions.

\end{document}